\documentclass{PoS}

\title{Recent progress in lattice supersymmetry: \\ from lattice gauge theory to black holes}

\ShortTitle{Recent progress in lattice supersymmetry: from lattice gauge theory to black holes}

\author{\speaker{Daisuke Kadoh}\\
Keio University, Hiyoshi 4-1-1, Yokohama, Kanagawa 223-8521, Japan\\
        E-mail: \email{%kadoh@post.kek.jp, 
        kadoh@keio.jp}}

\abstract{
Supersymmetry (SUSY) is a fascinating topic in theoretical physics, because of its unique and counterintuitive properties. 
It is expected to emerge as new physics beyond the standard model, 
and it is also a building block for supergravity and superstring theory.
A number of exact results obtained via SUSY theories provide insights into field theory. 
However, the dynamics of many SUSY theories are not yet fully understood, and
numerical study of SUSY theories through lattice simulations is promising as regards furthering this understanding. 
In this paper, I overview the current status of lattice SUSY by discussing its development in chronological order, and by reviewing some simple models. In addition, I discuss the numerical verification of gauge/gravity duality, which is one of the recent significant developments in this field.}

\FullConference{The 33rd International Symposium on Lattice Field Theory\\
                 14 -18 July  2015\\
                 Kobe International Conference Center, Kobe, Japan}

\begin{document}

\section{Introduction}

%
%  Motivation of supersymmerty
%
Supersymmetry (SUSY) has been intensively studied in theoretical physics for a long period of time, motivated by a variety of reasons. 
For example, supergravity includes this concept as a local symmetry and yields a field theoretical description of gravity
with the improved ultraviolet behavior
 \cite{Sugra}. 
In addition, SUSY is also included in superstring theory which may produce a ``theory of everything'' \cite{String}. 
SUSY theory has been studied as a strong candidate for physics beyond the standard model \cite{SUSY_phenomenology}.   
Furthermore,  many exact results and predictions have been obtained in SUSY theories,
for instance,  
Seiberg duality\cite{Seiberg:1994pq}, 
Seiberg-Witten theory\cite{Seiberg_Witten}, %\cite{Seiberg:1994rs, Seiberg:1994aj}, 
AdS/CFT\cite{Maldacena:1997re}, 
and provide deep insights into various aspects of quantum field theory. 
Nevertheless,  it is important to reveal the dynamics of SUSY theories in more detail, as they are not yet fully understood.

%
%  Lattice supersymmetry
%  
%  ... lattice is powerful for both QCD and SUSY 
% 
Lattice field theory has been used to determine the dynamics of field theory via {\it ab-initio} calculations and has achieved great success in the form of lattice quantum chromodynamics (QCD) \cite{Wilson:1974sk, LatticeQCD}. Therefore, 
a natural extension is to apply the lattice theory to SUSY, in order to study the non-perturbative physics of the latter. 
However, SUSY algebra has the form \cite{Wess_Bagger} 
\begin{eqnarray}
\{Q_\alpha, \bar Q_{\dot\beta} \} = 2 \sigma^m_{\alpha \dot\beta} P_m,
\end{eqnarray}
contains the infinitesimal translation operator $P_m$ which is broken on the lattice.  
Thus, a naive lattice discretization breaks the SUSY at the cutoff scale.

Although broken SUSY is, of course, restored at the classical continuum limit, such restoration does not occur 
at the quantum level, because ultraviolet divergence yields the relevant SUSY breaking operators. 
The supersymmetric continuum limit is achieved at the quantum level by fine tuning the relevant SUSY breaking operators,
and $N=1$ supersymmetric Yang-Mills (SYM) has already been studied by using fine tuning the gluino mass 
\cite{Montvay_works, desy_munster_hp}.  
However, such a ``brute-force'' approach is ineffective for general SUSY theories  in practice, because this fine tuning is computationally expensive.  
Therefore, it is difficult to apply lattice theory to SUSY theory.

%
% Many attempt fo ....
%
A large number of attempts have been made to solve this problem. 
One possible solution is a partial realization of SUSY that yields a supersymmetric continuum limit without (or with reduced) fine tuning. 
In previous studies \cite{Sakai:1983dg, Cohen:2003xe, Sugino:2003yb, Catterall:2009it}, 
it has been found that some nilpotent supercharges can be realized on the lattice for theories with extended SUSY; 
for instance, the two-dimensional $N=2$ Wess-Zumino (WZ) model and gauge theories with extended SUSY. 
These lattice formulations have been constructed with a small number of exact supersymmetries and have already been used to study interesting phenomena such as SUSY breaking \cite{Kanamori_Sugino_Suzuki, Catterall:2015tta} and gauge/gravity duality \cite{Catterall:2010fx}-\cite{Catterall:2014mha}, \cite{Kadoh:2015mka}.
Thus, lattice SUSY has become a feasible and powerful tool for revealing the non-perturbative physics of SUSY theories.

% 
% In this paper
%
In this paper, the current status of research into lattice SUSY is reviewed.
First, an overview of the history of lattice SUSY research is provided and the manner in which partial SUSY can be realized on the lattice in simple models is explained in section 2. 
Then, recent lattice results for numerical tests of gauge/gravity duality in maximally SYM theories are discussed in section 3. 
Section 4 contains a summary and a discussion of the future outlook for this field.

\section{Lattice supersymmetry}
The first paper on lattice SUSY was published in 1976, and many studies related to this topic have since been conducted. 
Approximately 24,000 papers have been published on the topic of lattice field theory, and approximately 400 papers of which discuss lattice 
SUSY.\footnote{Approximately 1,000 papers can be found in INSPIRE by searching for two keywords: ``lattice'' and ``supersymmetry''.
However, 600 of those papers are irrelevant to this topic. In fact, the number of relevant papers is approximately 250, 
excluding 130 $\sim$ 150 proceedings. }
In this section, the history of lattice SUSY is summarized and the manner in which partial SUSY can be realized on a lattice via super quantum mechanics and SYM is reviewed.

\subsection{History of lattice supersymmetry research}
SUSY was introduced into field theory at the beginning of the 1970s \cite{Gervais:1971ji},
lattice field theory was first proposed in the 1960s, and lattice gauge theory was first formulated in 1974 \cite{Wilson:1974sk}.
The first paper on lattice SUSY was published in 1976 by Dondi and Nicolai, who combined the above frameworks \cite{Dondi:1976tx}.
To obtain SUSY invariance of the action, the Leibniz rule 
\begin{eqnarray}
\partial_\mu (f g) = (\partial_\mu f) g +f (\partial_\mu g),
\label{LR}
\end{eqnarray}
is required. However, this rule is broken when the derivative $\partial_\mu$ is replaced with a local difference operator. 
Dondi and Nicolai \cite{Dondi:1976tx} noted that this breaking prevents the realization of SUSY on the lattice and defined the lattice WZ model without locality. 
No other early studies 
%\cite{Banks:1982ut, Elitzur:1982vh, Ichinose:1982ug, Bartels:1982ue, Nojiri:1984ys} 
obtained lattice formulations with exact SUSY and locality \cite{Banks:1982ut}-\cite{Nojiri:1984ys}.

The first lattice action with exact SUSY was reported by Sakai and Sakamoto in 1983 \cite{Sakai:1983dg}, for a two-dimensional $N=2$ WZ model
and using the Wilson fermion. 
\footnote{ Cecotti and Girardello independently discovered the same method using a Hamiltonian formulation\cite{Cecotti:1982ad}. }
This approach was based on a Nicolai-Parisi-Sourlas transformation %\cite{Nicolai:1979nr, Nicolai:1980jc,Parisi:1982ud} 
\cite{Nicolai_map, Parisi:1982ud}  
(so-called ``Nicolai mapping''), through which the boson action becomes a simple Gaussian integral and  the Jacobian of the transformation gives the fermion determinant.
The theory has an exact nilpotent supercharge $Q$, and the action is $Q$-exact:
\begin{eqnarray}
S= QV, \qquad Q^2=0.
\end{eqnarray}
Thus, the lattice action is invariant under $Q$-transformation without the Leibniz rule.
In section 2.2, this method is explained from the perspective of $Q$-exact action in SUSY quantum mechanics, which is another example involving Nicolai mapping \cite{Catterall:2000rv, Giedt:2004vb}.  
Lattice formulation with Nicolai mapping, refinements to this approach, and the other formulations of the WZ model have been studied intensively
%\cite{Bartels:1983wm, Elitzur:1983nj, Golterman:1988ta, Bietenholz:1998qq, Fujikawa:2001ka, Fujikawa:2002ic, Catterall:2001fr, Fujikawa:2002pa,  Kikukawa:2002as, Bonini:2004pm, Giedt:2004qs, Kikukawa:2004dd, Kadoh:2010ca}, 
\cite{Bartels:1983wm}-\cite{Kadoh:2010ca}, 
and lattice WZ models have also  been numerically examined %\cite{Beccaria:1998vi, Catterall:2003ae, Bergner:2007pu, Kastner:2008zc, Yu:2010zv}.  
\cite{Beccaria:1998vi}-\cite{ Yu:2010zv}.  
The sigma model has also been formulated on the lattice 
%\cite{ Catterall:2003uf, Catterall:2006sj, Flore:2012xj}.
\cite{Catterall_:Ghadab, Flore:2012xj}.

In the 1990s, lattice SUSY progressed to $N=1$ SYM in four dimensions, 
which consists of a gauge field (gluon) and a massless Majorana fermion (gluino).
Montvay began a numerical simulation of $N=1$ $SU(2)$ SYM with the Wilson fermion in the mid 1990s \cite{Montvay_works}.  
The only relevant SUSY breaking operator is the gluino mass term, and one can construct the massless limit that corresponds to the SUSY limit by fine-tuning the gluino mass \cite{Curci:1986sm,Suzuki:2012pc}.
Montvay's approach was then employed by the DESY-Munster collaboration,  and the resultant series of work is the largest in the field of lattice SUSY from 1996 onwards \cite{desy_munster_hp}. 
In recent papers \cite{desy_munster_recent}, %\cite{Bergner:2015lba, Bergner:2015adz},
 this group has observed the mass degeneracy between bosonic and fermionic bound states at low energy, which was theoretically predicted in \cite{Veneziano:1982ah}.  

%$N=1$ SYM was also numerically studied using Wilson fermion by the other groups\cite{Donini:1997hh}. 

Other significant progress in the 1990s is directly related to the discovery of the domain wall fermion 
%\cite{Kaplan:1992bt, Shamir:1993zy, Furman:1994ky} 
\cite{Kaplan:1992bt}-\cite{Furman:1994ky} 
and the overlap fermion
%\cite{Neuberger:1997fp, Neuberger:1998wv} 
\cite{Neuberger_works} 
obeying the Ginsparg-Wilson relation \cite{Ginsparg:1981bj}. 
The overlap fermion realizes the exact chiral symmetry \cite{Luscher:1998pqa} beyond the constraint of the Nielsen-Ninomiya no-go theorem \cite{Nielsen_Ninomiya}, 
%\cite{Nielsen:1980rz, Nielsen:1981xu}, 
and defines the massless fermions on the lattice. 
If a massless gauge field and a massless adjoint fermion exist, they form a supersymmetric pair. In other words,  
the exact chiral symmetry forbids the gluino mass and  realizes $N=1$ SYM in four dimensions without fine tuning. 
In this sense, lattice formulations without fine-tuning are discussed in 
%\cite{Nishimura:1997vg, Maru:1997kh, Neuberger:1997bg,Aoyama:1998in, Kaplan:1999jn}.
\cite{Nishimura_works}-\cite{Kaplan:1999jn}.
Using the domain wall fermion, $N=1$ SYM has been numerically investigated 
%\cite{Fleming:2000fa,Giedt:2008xm, Endres:2009yp}.
  \cite{Fleming:2000fa}-\cite{Endres:2009yp}.

As regards developments in the 21st century, significant progress was made in relation to SUSY gauge theory in 2003. 
Lattice formulations of the extended SUSY gauge theories were discovered by Cohen et al.\ \cite{Cohen:2003xe}, 
and independently by Sugino \cite{Sugino:2003yb}. 
The corresponding actions have exact partial SUSY on the lattice. 
In perturbation theory, the power counting theorem states that exact SUSY yields the correct continuum limit with no fine tuning in two dimensions, and with less fine tuning in three and four dimensions. 
As explained below, many variants have been discussed to date and other formulations have also been proposed. For instance,  Dadda et al. have constructed lattice actions by respecting SUSY algebra (the link approach) 
%\cite{D'Adda:2004jb, D'Adda:2005zk, D'Adda:2007ax}, 
\cite{link_approach}, 
and Catterall has defined lattice actions with partial SUSY  (the geometrical approach) 
%\cite{Catterall:2003wd, Catterall:2004np, Catterall:2005fd, Catterall:2006jw, Catterall:2006is}.
\cite{Catterall_geometrical_approach}.

Cohen et al.\ have constructed lattice SYM theories using orbifolding from the matrix models obtained by reducing the dimensions of the target theories.
The orbifolding procedure does not commute with all of SUSY, and the resultant lattice actions have partial SUSY.
The lattice actions for SYM with four and eight supercharges were given in \cite{Cohen:2003xe} and \cite{Cohen:2003qw}, respectively, 
after which Kaplan and Unsal reported a lattice SYM with sixteen supercharges \cite{Kaplan:2005ta, Unsal:2008kx}. 
The original Cohen-Kaplan-Katz-Unsal (CKKU) method uses a non-compact gauge field, 
and Unsal has shown that the compact gauge field can also be used \cite{Unsal:2005yh}. 
The lattice gauge theories using orbifolding are classified in \cite{Damgaard:2007be}.
Moreover, SUSY gauge theories with matter fields have been formulated on the lattice by Endres and Kaplan \cite{Endres:2006ic} 
for adjoint matter and by Matsuura \cite{Matsuura:2008cfa} for fundamental matter.
In addition, Joseph has utilized the CKKU method to define several theories
%\cite{Joseph:2013jya, Joseph:2013bra, Joseph:2014bwa}, 
\cite{Joseph_works}, 
and  Giedt has discussed the positivity or non-positivity of the fermion determinant for several related theories
%\cite{Giedt:2003ve, Giedt:2003vy, Giedt:2006pd}. 
\cite{Giedt_works}. 
The CKKU model has already been studied 
%perturbatively\cite{Onogi:2005cz} and 
numerically \cite{Catterall:2008dv}.

Sugino has defined a lattice action with a partial SUSY $Q$ that is nilpotent up to a gauge transformation,  by focusing on the $Q$-exactness of the SYM action, for several SYM theories %\cite{Sugino:2003yb, Sugino:2004qd, Sugino:2004uv, Sugino:2006uf, Sugino:2008yp}.
\cite{Sugino:2003yb}-\cite{Sugino:2006uf}. This method is based on the topological field theory \cite{TFT}, and  the $Q$-invariance of the action comes from its algebraic property and the symmetry can be kept even on the lattice, as explained in section 2.3.  
Later, Sugino defined a lattice action of two-dimensional $N=2$ supersymmetric QCD (SQCD) with an equal number of fundamental and anti-fundamental fermions \cite{Sugino:2008yp}, while the same theory was also formulated on the lattice by
Kikukawa and Sugino using the Ginsparg-Wilson relation for flavor symmetry to avoid the extra restriction on the number of fermions \cite{Kikukawa:2008xw}, and by Sugino, Suzuki, and the present author using a different type of twist and the Wilson fermion \cite{Kadoh:2009yf}.
 The importance of the admissible gauge action in Sugino's method was first noted in \cite{Sugino:2004qd}. 
Moreover, Matsuura and Sugino have proposed another type of gauge action without an admissibility condition \cite{Matsuura:2014pua}.
The Sugino model has already been studied numerically 
%\cite{Suzuki:2007jt, Kanamori:2008bk, Kanamori:2008yy,Hanada:2009hq,Giguere:2015cga}.
\cite{Kanamori_Suzuki}-\cite{Hanada:2009hq}, \cite{Giguere:2015cga}.
In addition, numerical evaluation of the SUSY Ward-Takahashi identity \cite{Kanamori_Suzuki, Giguere:2015cga} indicates that the Sugino model reproduces a desirable continuum theory in two dimensions, beyond the perturbation theory.

The lattice formulations discussed above are similar in the sense that the remnant charges are nilpotent at least up to gauge transformations and the lattice actions are written in $Q$-exact form although they are based on different methods \cite{Unsal:2006qp}.   
If fact, a certain equivalence and similarity between these formulations has been discussed in several papers.
%\cite{Takimi:2007nn, Damgaard:2007xi, Damgaard:2007eh, Catterall:2007kn, Damgaard:2008pa}.  
In \cite{Takimi:2007nn}, it is shown that the Sugino model can be reproduced from the Catterall geometrical approach, by truncating the degrees of freedom of the complexified fields while preserving the supercharge. 
Further, Damgaard and Matsuura have shown that the Catterall geometrical approach is equivalent to the CKKU method \cite{Damgaard_Matsuura} \footnote{Such equivalence is also discussed in 
\cite{Catterall:2007kn}.},  and have also shown that the lattice action given by the link approach is equivalent to  the CKKU action  \cite{Damgaard:2007eh}.

Since the discovery of lattice SUSY actions with a few exact supersymmetries,  many numerical applications have been conducted. 
For example, Kanamori, Suzuki and Sugino have studied two-dimensional $N=2$ SYM using the Sugino action and  measured the vacuum expectation value of the Hamiltonian defined from  exact $Q$-symmetry; hence they examined  whether or not SUSY is broken in that theory
\cite{Kanamori_Sugino_Suzuki}. Further, Catterall and Veernala studied the two-dimensional $N=2$ SQCD from a similar perspective \cite{Catterall:2015tta}. For non-gauge theory, SUSY breaking  has been studied in 
%\cite{Beccaria:2004pa, Wozar:2011gu, Steinhauer:2014yaa}. 
\cite{Beccaria:2004pa}-\cite{Steinhauer:2014yaa}. 
 In addition, Kikukawa and Kawai have examined the two-dimensional $N=2$ WZ model and shown that this theory reproduces $N=2$ superconformal field theory %for a cubic superpotential 
 \cite{Kawai:2010yj}.

Remarkable applications include numerical verifications of gauge gravity duality in maximally SYM. 
In one dimension, Catterall and Wiseman have employed a naive lattice action without exact SUSY 
%\cite{Catterall:2007fp, Catterall:2008yz, Catterall:2009xn} 
to study the thermodynamics of a dual black hole \cite{Catterall_Wiseman}.
\footnote{Hanada et al. have studied the same theory numerically using a sharp momentum cut-off \cite{Hanada_Nishimura}. } 
Numerical simulation with the Sugino lattice action has also been performed and the validity of the gauge/gravity duality in this system at next-to-leading order in a low temperature expansion has been shown \cite{Kadoh:2015mka}.  In two dimensions, Catterall, Wiseman and Joseph have employed the CKKU lattice action to study a black hole - black string phase transition \cite{Catterall:2010fx}. Giguere and the present author have employed the Sugino lattice action to examine the duality in two dimensions \cite{Giguere:2015cga,Giguere:2016}, 
which will be introduced in section 3.
Further, Catterall et al.\ have examined four-dimensional $N=4$ SYM from a lattice SYM and its simulations, and tried to verify AdS/CFT 
%\cite{Catterall:2011pd,  Catterall:2012yq, Catterall:2013roa, Catterall:2014vka, Catterall:2014mha}. 
\cite{Catterall:2011pd}-\cite{Catterall:2014mha}. 
%Four dimensional $N=4$ SYM are given from two dimensional CKKU model\cite{Hanada:2010kt} and from two dimensional Sugino model on fuzzy two sphere\cite{Hanada:2011qx, Matsuura:2015caa}.

There are many interesting works that should also be introduced here; for instance, 
the energy-momentum tensor in $N=1$ SYM \cite{Suzuki:2012wx, Suzuki:2013gi}, 
%gradient flow in lattice SUSY gauge theory\cite{Kikuchi:2014rla}, 
the cyclic Leibniz rule \cite{Kato:2013sba, Kadoh:2015zza}, 
and topological  twisted SYM on a discretized Riemann surface \cite{Matsuura_Misumi_Ohta}. 
%loop formulation\cite{Steinhauer:2014oda, Baumgartner:2014nka, Baumgartner:2015qba, Baumgartner:2015zna}, 
%other formulations \cite{Harada:2003bs, Suzuki:2005dx, Elliott:2005bd, Hanada:2010gs} 
Unfortunately, introducing and evaluating these studies would extend this review beyond the page limit, therefore the author would like to leave them to another reviewer.

\subsection{Supersymmetric quantum mechanics}
In this section, lattice $N=2$ SUSY quantum mechanics (QM) is reviewed, which provides a clear concept of the $Q$-exact formulation of lattice SUSY.  

The continuum euclidean action of SUSY QM is given by 
\begin{eqnarray}
S_{\rm SQM}
 =
 \int dt \left\{
   \frac{1}{2}\left(\frac{d\phi}{dt}\right)^2+\frac{1}{2}F^2+iFW(\phi)
   +i\bar{\psi}\left(\frac{d}{dt}+\frac{\partial W(\phi)}{\partial\phi}\right)\psi
 \right\},
 \label{SQM_con}
\end{eqnarray}
where $\phi(t)$ is a real scalar field, $F(t)$ is a real auxiliary field, 
and $\psi(t)$ and $\bar\psi(t)$ are one-component fermions  with euclidean time $t$. The off-shell SUSY transformation with two Grassmann parameters $\epsilon$ and $\bar \epsilon$,
\begin{eqnarray}
\label{SUSY_con_s}
 \delta\phi &=& \epsilon\bar\psi-\bar\epsilon\psi,\\
 \delta\psi &=& \epsilon\left(i\frac{d\phi}{dt}+F\right), \qquad 
 \delta\bar\psi =\bar\epsilon\left(-i\frac{d\phi}{dt}+F\right),\\ 
\label{SUSY_con_e}
 \delta F &=& -i\epsilon\frac{d\bar\psi}{dt}-i\bar\epsilon\frac{d\psi}{dt},
\end{eqnarray}
indicates the invariance of the action (\ref{SQM_con}) for any function $W(\phi)$, 
as the Leibniz rule (\ref{LR}) states that the Lagrangian varies up to a total divergence, $\delta L=dX/dt$
with $X=\epsilon \bar\psi \left(\frac{d\phi}{dt} -iF +W \right)+\bar \epsilon \psi W$, and $\delta S=0$. 

One can easily define a naive lattice action by placing the fields on the lattice sites and  replacing the $t$-derivative in (\ref{SQM_con})-(\ref{SUSY_con_e}) and the integral in  (\ref{SQM_con})  with a  difference operator and summation over the sites, respectively. 
However, this naive action breaks SUSY at a finite lattice spacing because 
the Leibniz rule does not hold for any difference operators \cite{Dondi:1976tx, Kato:2008sp}.

Instead, the $Q$-exact form of the action allows either $\epsilon$ or $\bar \epsilon$ invariance to be realized on the lattice.
$Q$ and $\bar Q$ are supertransformations extracted from (\ref{SUSY_con_s})-(\ref{SUSY_con_e}), with  $\delta=\epsilon \bar Q +\bar\epsilon Q$, and satisfy  
\begin{eqnarray}
Q^2=0,\quad \bar Q^2=0, \quad \{Q,\bar Q \}=-2i\frac{d}{dt}.
\end{eqnarray}
The action given in (\ref{SQM_con}) is then expressed in  $Q$-exact form as
\begin{eqnarray}
S_{\rm SQM}
 = Q
 \int dt \left\{ \frac{1}{2} \bar\psi \left(
   F+ i \frac{d\phi}{dt} + 2 i W(\phi)\right)
 \right\}.
 \label{SQM_Qexact}
\end{eqnarray}
  We can show that the $Q$-exact action (\ref{SQM_Qexact}) is $Q$-invariant without the Leibniz rule, as $Q^2=0$.  
  In other words, we have already employed the Leibniz rule to express the action in the form given in (\ref{SQM_Qexact}).
  Although $\bar Q$ invariance requires the Leibniz rule because the anti-commutator of $Q$ and $\bar Q$ is a derivative, the $Q$-invariance does not require the Leibniz rule and can be realized on the lattice. 

The $Q$-exact action (\ref{SQM_Qexact}) leads to the lattice action,
\begin{eqnarray}
S_{\rm LAT}
 = Q \, a
  \sum_t \left\{ \frac{1}{2} \bar\psi \left(
   F+ i \Delta \phi + 2 i W(\phi)\right)
 \right\},
 \label{SQM_Qexact_lat}
\end{eqnarray}
where $\Delta$ is a forward difference operator, with
%\begin{eqnarray}
$\Delta \phi(t)= \frac{\phi(t+a) -\phi(t)}{a}$.
%\end{eqnarray}
The lattice Q-transformation 
\begin{eqnarray}
\label{Q_lat}
&&
 Q\phi = -\psi,\qquad
 Q\psi =0, \qquad 
 Q\bar\psi =F-i \Delta \phi,\qquad
 Q F = -i \Delta\psi,
\end{eqnarray}
satisfies $Q^2=0$, even for finite lattice spacing. Therefore, the lattice action (\ref{SQM_Qexact_lat}) is manifestly invariant under partial supersymmetry $Q$ on the  lattice. 

It is obvious that (\ref{SQM_Qexact_lat})  coincides with the continuum action given in (\ref{SQM_con}) via (\ref{SQM_Qexact}) for the classical continuum limit. 
Numerical simulations with the lattice SUSY action given in (\ref{SQM_Qexact_lat}) also indicate that this formalism reproduces SUSY QM beyond perturbation theory \cite{Catterall:2000rv, Giedt:2004vb}.

\subsection{Supersymmetric Yang-Mills theory}
A definition of lattice SUSY QM was presented in the previous section and the manner in which partial supersymmetry $Q$ can be maintained on the lattice as an exact symmetry was explained.
This approach is based on $Q$ being nilpotent and $Q$-exactness of the action. The same method is applicable to the gauge theory. In this section, the Sugino action of two-dimensional SYM with four supercharges is explained \cite{Sugino:2003yb}.

In two dimensions, SYM with four supercharges has a gauge field $A_\mu$, a complex scalar $\phi$,  a Dirac fermion $\psi=(\psi_L,\psi_R)^T$ and  an auxiliary field $D$
as field variables. Its euclidean action is defined by
\begin{eqnarray}
&& S_{\rm SYM}= \frac{1}{g^2} \int {\rm d}^2 x\ {\rm tr} \bigg\{
   \frac{1}{2} F_{\mu\nu} F_{\mu\nu} 
  + D_\mu \phi D_\mu \bar\phi
  + \frac{1}{4} [\phi, \bar\phi]^2
  +D^2
\nonumber \\
&& \hspace{1cm} 
  + 4 \bar \psi_R (D_1 - i D_2) \psi_R
  + 4 \bar \psi_L (D_1 + i D_2) \psi_L
  + 2 \bar \psi_R [\bar\phi,  \psi_L]
  + 2 \bar \psi_L [\phi, \psi_R]
\bigg\},
\label{sym_action}
\end{eqnarray}
where $\mu,\nu=1,2$ and 
\begin{eqnarray}
F_{\mu\nu} &=&\partial_\mu A_\nu - \partial_\nu A_\mu +i [A_\mu,A_\nu],\label{cont_field_tensor}\\
D_\mu \phi &=& \partial_\mu \phi +i[A_\mu,\phi].
\end{eqnarray}
The infinitesimal gauge transformation, 
\begin{eqnarray}
\delta_\omega A_\mu = - D_\mu \omega, \qquad \delta_\omega f = -i [f,\omega], \quad  (f=\phi, \psi,D),
\label{gauge_transf}
\end{eqnarray}
makes the action given in (\ref{sym_action}) invariant for any gauge parameter $\omega$. Although the $\omega^a$ are originally real, the invariance holds for complexified parameters $\omega^a \in \mathbb{C}$.

We can express (\ref{sym_action}) in $Q$-exact form using a nilpotent $Q$ as in the case of SUSY QM.  To achieve this easily,  
topological field theory (TFT) field variables can be employed \cite{TFT}. \footnote{ In TFT  \cite{TFT},  the Lorentz group is redefined by a  topological twist and the fields transform in a new manner
under the twisted Lorentz group; for instance, the fermion transforms like a vector or scalar. In this notation, a nilpotent scalar supercharge $Q$ is identified with $Q \equiv  -(Q_L + \bar Q_R)/\sqrt{2}$ and the twisted fermion fields $\psi_\mu,\chi, \eta$ are defined by
\begin{eqnarray}
\psi_1=\frac{1}{\sqrt{2}}  (\psi_L+\bar\psi_R), \quad \psi_2=\frac{i}{\sqrt{2}}  (\psi_L-\bar\psi_R),\quad 
\chi = \frac{1}{\sqrt{2}}  (\psi_R-\bar\psi_L)  ,\qquad \eta = -i \sqrt{2}  (\psi_R+\bar\psi_L).
\end{eqnarray}
See the appendix of \cite{Sugino:2008yp} for more details.
}
%In TFT,  a nilpotent scalar supercharge $Q$ defines physical quantities that are restricted in the $Q$-closed form.
%Then, the action can be expressed in $Q$-exact form and  the transformation laws under $Q$ become slightly simpler than those of the original variables. Without restricting the physical quantities,  we may employ twisted fields to express the original theory, because the twisting procedure does not affect the theory on a flat space. 
%
%
The action given in (\ref{sym_action}) is expressed as the following simple $Q$-exact form in twisted fields: 
\begin{eqnarray}
S_{cont}=Q \frac{1}{2g^2} \int {\rm d}^2 x \  {\rm tr} \bigg\{
\chi (H -2i F_{12}) + \frac{1}{4} \eta [\phi,\bar \phi] -i \psi_\mu D_\mu\bar\phi
  \bigg\},
  \label{cont_SYM_Qexact}
\end{eqnarray}
where $Q$ acts on the fields as
\begin{eqnarray}
&& Q A_\mu = \psi_\mu, \qquad Q\psi_\mu = i D_\mu \phi, \\
&& Q \phi=0,  \label{cont_Q_phi}\\
&& Q \chi  = H, \qquad QH= [\phi,\chi],\\
&& Q \bar \phi= \eta,\quad \hspace{5mm}  Q\eta=[\phi,\bar\phi],   \label{cont_Q_eta}
\end{eqnarray}
and satisfy $Q^2= -i \delta_\phi$ (the gauge transformation with the parameter $\phi$).

We define the lattice theory on a two dimensional lattice with a lattice spacing $a$, and set $a=1$ for simplicity. 
The fermion, scalar and auxiliary fields are defined on the sites, while  the gauge fields are defined on the links as link fields $U_\mu(x) \in G$.
Under the lattice gauge transformations, the link fields transform as
\begin{eqnarray}
\delta_\omega U_\mu(x) = i \omega(x) U_\mu(x) - iU_\mu(x) \omega(x+\hat\mu),
\end{eqnarray}
and the fermion, scalar and auxiliary fields transform like the second equation of (\ref{gauge_transf}).
The covariant forward difference operator  $\nabla_\mu$  and  the plaquette $P_{\mu\nu}(x)$ defined by
\begin{eqnarray}
&& \nabla_\mu \phi(x) = U_\mu(x)\phi(x+\hat \mu) U^{-1}_\mu(x) - \phi(x), \label{fwd_D}\\
&& P_{\mu\nu}(x) = U_\mu(x) U_\nu(x+\hat \mu) U^\dag_\mu(x+\hat \nu) U^\dag_\nu(x), 
\end{eqnarray}
are gauge covariant. 

We can formally define the Sugino lattice action from the $Q$-exact action (\ref{cont_SYM_Qexact}) by replacing the integral over the space time with summation over the lattice sites, and by replacing the covariant derivative and the field strength with their lattice respective versions, such that
\begin{eqnarray}
S_{LAT}=Q \frac{1}{2g^2} \sum_{x} \  {\rm tr} \bigg\{
\chi (H -2i F^{lat}_{12}) + \frac{1}{4} \eta [\phi,\bar \phi] -i \psi_\mu \nabla_\mu\bar\phi
  \bigg\},
  \label{sugino_action}
\end{eqnarray} 
where $\nabla_\mu$ is the covariant forward difference operator (\ref{fwd_D}) and $F^{lat}_{12}$ is a lattice version of (\ref{cont_field_tensor}). 
There are a number of possible choices for $F^{lat}_{12}$, but a simple choice $F^{lat}_{12} =-\frac{i}{2}(P_{12}-P_{12}^\dag)$ leads to degenerate vacua that are irrelevant to the correct continuum limit.  The admissible field tensor,
\begin{eqnarray}
F^{lat}_{12} (x)
\equiv -\frac{i}{2} {(P_{12}-P_{12}^\dag)} \left\{1-\frac{1}{\epsilon^2 }\vert \vert 1- P_{12}(x) \vert\vert \right\}^{-1},
\end{eqnarray}
avoids the  extra vacua for sufficiently small $\epsilon$ \cite{Sugino:2004qd}. Moreover, instead of the admissible field tensor, a ${\rm tan}(\theta/2)$-type tensor can also be used \cite{Matsuura:2014pua}.

The lattice $Q$-transformations of $U_\mu$ and $\psi_\mu$ are
\begin{eqnarray}
 Q U_\mu(x) =i \psi_\mu(x) U_\mu(x), \qquad Q \psi_\mu(x) = i \nabla_\mu \phi (x) + i \psi_\mu(x) \psi_\mu(x),
%&& Q \phi=0,  \\
%&& Q \chi  = H, \qquad QH= [\phi,\chi],\\
%&& Q \bar \phi= \eta,\quad \hspace{5mm}  Q\eta=[\phi,\bar\phi],  
\end{eqnarray}
while the other transformations are identical to the continuum transformations, (\ref{cont_Q_phi})-(\ref{cont_Q_eta}).
It is easy to show that $Q^2=-i\delta_\phi $ even on the lattice. Thus, $Q$ is an exact symmetry of the lattice action give in (\ref{sugino_action}).

In the naive continuum limit, the Sugino action given in (\ref{sugino_action}) reproduces (\ref{cont_SYM_Qexact}), that is, the action given in (\ref{sym_action}). In two dimensions, perturbative power counting indicates that the only relevant SUSY breaking operator is the mass term for the scalar field.
The exact $Q$-symmetry and $U(1)_R$ symmetry of the Sugino action forbid the mass term, and the SUSY is restored 
in the quantum continuum limit (in  perturbation theory, at least) \cite{Sugino:2003yb}. 
The numerical result for the SUSY Ward-Takahashi identity indicates that the Sugino action reproduces the desirable continuum theory beyond the perturbation theory at the continuum limit in two dimensions \cite{Kanamori_Suzuki, Giguere:2015cga}.

\section{Gauge/gravity duality and lattice gauge theory}
In this section, the present author's lattice studies that show the validity of the gauge/gravity duality of maximally SYM in one and two dimensions are presented \cite{Kadoh:2015mka, Giguere:2015cga, Giguere:2016}.

The gauge/gravity duality, such as AdS/CFT, states that gauge theory is equivalent to the theory of gravity on the basis of superstring theory. 
Therefore, the duality can be employed to study strongly coupled gauge theory using the corresponding gravitational theory, and conversely, to study the quantum behavior of gravity from the gauge theory perspective. 
Moreover, this possibility implies that a gauge theory yields a definition of superstring theory beyond its perturbative description.  
In this sense, many interesting applications arise once the concept of gauge/gravity duality is accepted.  
However, this duality has not yet been proven mathematically. Thus, it is important to test the validity of this duality.

Lattice gauge theory enables us to test gauge/gravity duality, as it is applicable to a strongly coupled gauge theory 
that describes dual classical gravity well.
We can examine whether or not this duality is accurate by comparing the lattice results in gauge theory with the theoretical predictions yielded by gravity theory. 
For finite temperature and low dimensional versions of AdS/CFT, 
$p+1$-dimensional maximally SYM is expected to be dual to black $p$-branes at low temperature for the large $N$ limit. As reported in previous works 
using a sharp momentum cut-off \cite{Hanada_Nishimura}-\cite{Filev:2015hia}, 
%\cite{Anagnostopoulos:2007fw, Hanada:2008gy, Hanada:2008ez, Hanada:2011fq, Hanada:2013rga} and lattice regularization 
%\cite{Catterall:2007fp, Catterall:2008yz, Catterall:2009xn, Kadoh:2015mka, Filev:2015hia}, 
%\cite{Catterall_Wiseman}-\cite{Filev:2015hia}, 
the one-dimensional theory reproduces black hole thermodynamics at low temperatures. 
For the two-dimensional case, the black hole-black string phase transition has been investigated \cite{Catterall:2010fx}.
Moreover, N=4 SYM has already been studied using lattice theory
%\cite{Catterall:2011pd,  Catterall:2012yq, Catterall:2013roa, Catterall:2014vka, Catterall:2014mha} 
 \cite{Catterall:2011pd}-\cite{Catterall:2014mha} 
in order to verify AdS/CFT.
Hereafter, the present author's lattice results given in \cite{Kadoh:2015mka} for $p=0$ and in \cite{Giguere:2015cga, Giguere:2016} for $p=1$ are considered.

The gravitational theory corresponding to one-dimensional maximally SYM predicts that the internal energy of the black hole behaves according to
\begin{eqnarray}
\frac{1}{N^2}
\left( \frac{E}{\lambda^{1/3}} \right)=c_1 
\left( \frac{T}{\lambda^{1/3}}
\right)^{2.8} 
+c_2 \left( \frac{T}{\lambda^{1/3}} 
\right)^{4.6}+ \cdots,
\label{energy_nlo}
\end{eqnarray}
at low temperature $T/\lambda^{1/3} \ll 1$
for the large $N$ limit. The value of the $c_1$ coefficient can be estimated from a type IIA ten-dimensional  SUGRA action\cite{Klebanov:1996un}:
\begin{eqnarray}
c_1=
\frac{9}{14} \left(4^{13} 15^2 \left(\frac{\pi}{7}\right)^{14}\right)^{\frac{1}{5}} 
=
7.407....
\end{eqnarray}
Here $c_2$ corresponds to  $\alpha'$ corrections in string theory, and its value is currently unknown.

From the gauge theory perspective, the internal energy can be given by the expectation value of the action itself:
\begin{eqnarray}
E =-\frac{3}{\beta} \langle S
\rangle,
\label{internal_energy_in_gauge_side1}
\end{eqnarray}
where $S$ is the $Q$-exact action of maximally SYM in one dimension and $\beta=1/T$.
The $Q$-exactness of the action yields the correct zero point of the internal energy (\ref{internal_energy_in_gauge_side1}) in the zero temperature limit corresponding to the SUSY limit. We integrate the fermions and auxiliary fields by hand and evaluate (\ref{internal_energy_in_gauge_side1}) from the boson action. \footnote{See also Ref. \cite{Catterall_Wiseman}.}

\begin{figure}[t]
 \begin{center}
  \includegraphics[width=70mm]{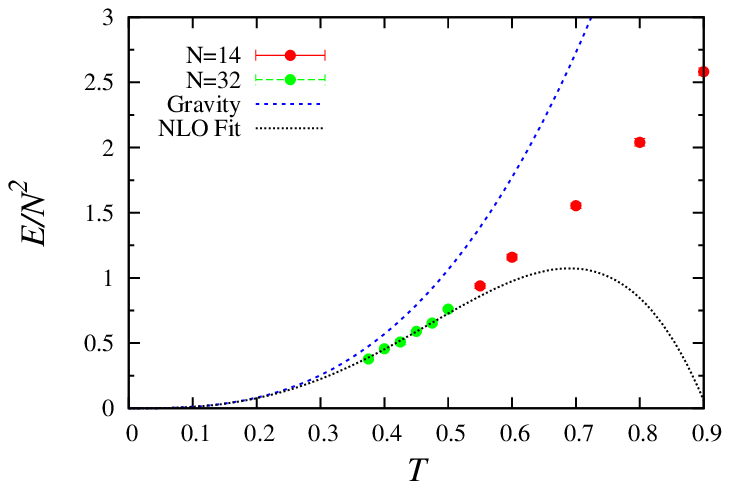}
  \includegraphics[width=70mm]{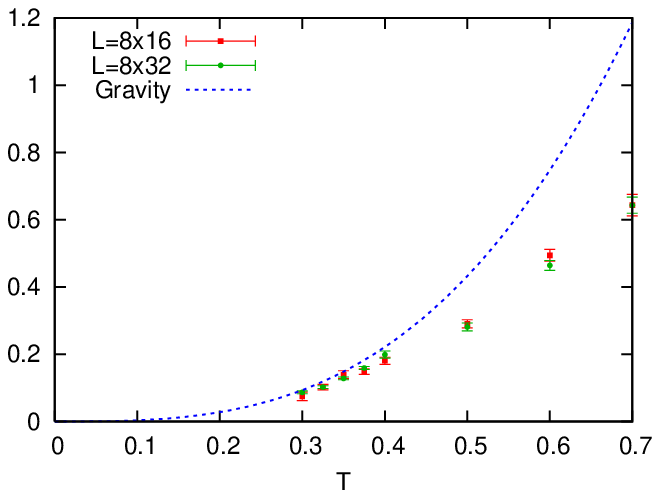}
   \caption{Thermodynamics of black $p$-branes. (Left) Internal energy of black hole ($p=0$ \cite{Kadoh:2015mka}), and (right) 
   $( \epsilon -p )/N^2$ for black string  ($p=1$ \cite{Giguere:2016}).  The dashed blue curves are the theoretical predictions 
   given by the gravitational theory  at leading order. 
   %The dashed curve in the left figure denotes the fit result obtained using the NLO formula.  
   The group size is fixed at $N=12$ in the right figure. 
   }
   \label{fig:internal_energy}
 \end{center}
\end{figure}

Figure \ref{fig:internal_energy} (left) shows the internal energy of the black hole.
In the lattice simulation of SYM in one dimension,  we employ the Sugino lattice action and quench the phase of the fermion pfaffian. 
The red and green points denote the results for $N=14$ and $32$, respectively. 
The $x$-axis denotes the dimensionless temperature $T_{\rm eff}=T/\lambda^{1/3}$. 
The lattice size is fixed at $L=16$ and the corresponding lattice spacing is $aT_{\rm eff}=1/L$.
The dashed blue curve represents the theoretical prediction yielded by the gravitational theory at the leading order of expansion, (\ref{energy_nlo}) with $c_2=0$.

The lattice data approach the theoretical prediction given by the gravitational theory 
as the temperature decreases. However, the temperatures employed in these simulations were not sufficiently low to explain the leading behavior.
%To obtain quantitative results for the leading-order term, simulations at further low temperatures are required.
Instead, one can extract the next-to-leading order (NLO) term from these results by fitting them using the formula
\begin{eqnarray}
f(x)= 7.41 x^{2.8} + C x^p,
\label{fit_formulta__nlo}
\end{eqnarray}
where $C$ and $p$ are the fitting parameters. 
From (\ref{energy_nlo}), 
if the assumption of duality is valid, the obtained $p$ should be $4.6$. 
We performed the fit using the five points within the $0.375 \le T \le 0.475$ range and obtained
\begin{eqnarray}
C=9.0 \pm 2.6,  \qquad p=4.74 \pm 0.35.
\label{fit_result_energy_nlo}
\end{eqnarray}

The obtained $p$ is consistent with the theoretical prediction within a statistical error of approximately 7\%. 
This is the first lattice result for the NLO term, which quantitatively shows the validity of the duality conjecture in this one dimensional theory \cite{Kadoh:2015mka}.

As an extension  to higher dimensions, we are performing a lattice simulation of maximally SYM in two dimensions with the Sugino lattice action.  In two dimensions, the mass term for the scalar field receives a quantum correction and becomes a relevant SUSY breaking operator. 
The exact SUSY renders the mass term irrelevant and guarantees the supersymmetric continuum limit in perturbation theory, at least.
The numerical results for the SUSY Ward-Takahashi identity show that SUSY is restored at the continuum limit beyond perturbation theory \cite{Giguere:2015cga}.

We consider this theory for finite temperature and on a compactified space on a circle with circumference $L$.
In this case, the expectation value of the action gives 
\begin{eqnarray}
\epsilon - p  =-\frac{2}{\beta L } \langle S
\rangle,
\label{internal_energy_in_gauge_side2}
\end{eqnarray}
where $\epsilon$ is the internal energy density and $p$ is pressure. The gravity dual of the target gauge theory is a black string for which 
\begin{eqnarray}
\epsilon - p = \frac{1}{3} c_0  N^2  T^3,
\label{e_minus_p_in_gravity_side}
\end{eqnarray}
with
%\begin{eqnarray}
$c_0 = \frac{2^4\pi^{5/2}}{3^3} (= 10.37...)$
%\end{eqnarray}
at low temperature and the large N limit.

Figure \ref{fig:internal_energy} (right) shows the internal energy minus pressure as a function of temperature, 
which was calculated via lattice simulations with the Sugino action.  
In those simulations,  $N=12$ was set and the phase of the fermion pfaffian was quenched.
It is apparent that, although the continuum limit has not yet been reached, the results reproduce the prediction well. 
This figure provides strong evidence of the duality between the target gauge theory and the dual gravity theory \cite{Giguere:2016}.

\section{Summary and future perspective}

Lattice SUSY is a powerful tool for revealing the non-perturbative physics of SUSY theories,  
which have been studied in a broad range of fields in theoretical physics. 
It is difficult to construct SUSY lattice theory because the Leibniz rule does not apply.
In this paper, the progress in the field of lattice SUSY has been reviewed, and lattice formulations that retain partial SUSY on the lattice
without  usage of the Leibniz rule have been emphasized.  
These formulations have already been used to study interesting aspects of SUSY theories, 
such as SUSY breaking, and the duality between maximally SYM and black branes.  
Many numerical studies have shown that these formulations are effective.

The realization of full SUSY on the lattice remains problematic. 
Some no-go theorems \cite{Kato:2008sp, Bergner:2009vg} suggest that it is difficult to retain full SUSY on a lattice with locality.  
This situation reminds us of the Nielsen-Ninomiya no-go theorem and the history of exact chiral symmetry. 
One would think that the important consequence of the Nielsen-Ninomiya theorem would be discouragement of any attempt to construct chiral invariant lattice models for QCD. 
However, exact chiral symmetry  has been achieved because of the discovery of the overlap fermion satisfying the Ginsparg-Wilson relation.  
Therefore efforts to construct fully SUSY invariant lattice models for SUSY theories should be continued.

Considerable future progress can be expected for lattice SUSY,  because it has a wide range of application. 
It is important that a method of obtaining four-dimensional lattice SUSY theories without fine tuning is developed. 
Four dimensional SYM can be obtained from two-dimensional lattice SYM \cite{Hanada_Matsuura_Sugino}, and 
numerical simulations with such formulations will provide us with a deep understanding of the interesting aspects of SUSY theories, for instance, Seiberg duality, Seiberg-Witten theory, and AdS/CFT.
Thus, further progress in lattice SUSY is expected in the near future.

\section*{Acknowledgment}
I would like to thank Noboru Kawamoto, Hiroshi Suzuki, Yoshio Kikukawa, Fumihiko Sugino, So Matsuura, Naoya Ukita,  Issaku Kanamori and Eric Giguere for many discussions about lattice supersymmetry and valuable comments on the draft, and Mitsuhiro Kato, Makoto Sakamoto,  and Hiroto So for their influence on my recent work concerning the cyclic Leibniz rule. I am thankful to the staff at the RIKEN RICC, HOKUSAI-GW, the K computer and the KEK supercomputer for their continued help and support.


\begin{thebibliography}{99}

%\cite{Sugra}
\bibitem{Sugra}
  D.~Z.~Freedman and  A.~Van Proeyen,
  \textit{Supergravity},
  Cambridge University Press (2012).

%\cite{String}
\bibitem{String}
M.B.~Green, J.H.~Schwarz and E.~Witten,
\textit{Superstring Theory},
Cambridge University Press (1987), \ \ \ J. Polchinski, 
\textit{String Theory I, II},
Cambridge University Press (1998).


%\cite{SUSY_phenomenology}
\bibitem{SUSY_phenomenology}
  M.~A.~Luty,
  \textit{2004 TASI lectures on supersymmetry breaking},
  hep-th/0509029. 
  %%CITATION = HEP-TH/0509029;%%
  %89 citations counted in INSPIRE as of 16 Mar 2016
  T.~Gherghetta,
    \textit{ TASI Lectures on a Holographic View of Beyond the Standard Model Physics,}
  arXiv:1008.2570 [hep-ph].
  %%CITATION = ARXIV:1008.2570;%%
  %42 citations counted in INSPIRE as of 13 mar 2015



%\cite{Seiberg:1994pq}
\bibitem{Seiberg:1994pq}
  N.~Seiberg,
  %``Electric - magnetic duality in supersymmetric nonAbelian gauge theories,''
  Nucl.\ Phys.\ B {\bf 435} (1995) 129.
  %doi:10.1016/0550-3213(94)00023-8
  %[hep-th/9411149].
  %%CITATION = doi:10.1016/0550-3213(94)00023-8;%%
  %1294 citations counted in INSPIRE as of 28 Feb 2016

%\cite{Seiberg_Witten}
\bibitem{Seiberg_Witten}
  N.~Seiberg and E.~Witten, 
  Nucl.\ Phys.\ B {\bf 426} (1994) 19
   [Nucl.\ Phys.\ B {\bf 430} (1994) 485], \ Nucl.\ Phys.\ B {\bf 431} (1994) 484.


%\cite{Seiberg:1994rs}
%\bibitem{Seiberg:1994rs}
%  N.~Seiberg and E.~Witten,
  %``Electric - magnetic duality, monopole condensation, and confinement in N=2 supersymmetric Yang-Mills theory,''
%  Nucl.\ Phys.\ B {\bf 426} (1994) 19
%   [Nucl.\ Phys.\ B {\bf 430} (1994) 485]
  %doi:10.1016/0550-3213(94)90124-4
  %[hep-th/9407087].
  %%CITATION = doi:10.1016/0550-3213(94)90124-4;%%
  %2722 citations counted in INSPIRE as of 28 Feb 2016
  
%\cite{Seiberg:1994aj}
%\bibitem{Seiberg:1994aj}
%  N.~Seiberg and E.~Witten,
  %``Monopoles, duality and chiral symmetry breaking in N=2 supersymmetric QCD,''
%  Nucl.\ Phys.\ B {\bf 431} (1994) 484
  %doi:10.1016/0550-3213(94)90214-3
  %[hep-th/9408099].
  %%CITATION = doi:10.1016/0550-3213(94)90214-3;%%
  %1974 citations counted in INSPIRE as of 28 Feb 2016


%\cite{Maldacena:1997re}
\bibitem{Maldacena:1997re}
  J.~M.~Maldacena,
  %``The Large N limit of superconformal field theories and supergravity,''
  Int.\ J.\ Theor.\ Phys.\  {\bf 38} (1999) 1113
   [Adv.\ Theor.\ Math.\ Phys.\  {\bf 2} (1998) 231].
  %doi:10.1023/A:1026654312961
  %[hep-th/9711200].
  %%CITATION = doi:10.1023/A:1026654312961;%%
  %11508 citations counted in INSPIRE as of 28 Feb 2016







%\cite{Wilson:1974sk}
\bibitem{Wilson:1974sk}
  K.~G.~Wilson,
  %``Confinement of Quarks,''
  Phys.\ Rev.\ D {\bf 10} (1974) 2445.
  %doi:10.1103/PhysRevD.10.2445
  %%CITATION = doi:10.1103/PhysRevD.10.2445;%%
  %4119 citations counted in INSPIRE as of 28 Feb 2016
  


%\cite{LatticeQCD}
\bibitem{LatticeQCD}
The 33rd International Symposium on Lattice Field Theory (LATTICE2015), http://www.aics.riken.jp/sympo/lattice2015/


%\cite{Wess_Bagger}
\bibitem{Wess_Bagger}
J. Wess and J. Bagger, \textit{Supersymmetry And Supergravity}, Princeton University
Press (1992).



%\cite{Gervais:1971ji}
\bibitem{Gervais:1971ji}
  J.~L.~Gervais and B.~Sakita,
  %``Field Theory Interpretation of Supergauges in Dual Models,''
  Nucl.\ Phys.\ B {\bf 34} (1971) 632.
  %doi:10.1016/0550-3213(71)90351-8
  %%CITATION = doi:10.1016/0550-3213(71)90351-8;%%
  %343 citations counted in INSPIRE as of 28 Feb 2016





%\cite{Dondi:1976tx}
\bibitem{Dondi:1976tx}
  P.~H.~Dondi and H.~Nicolai,
  %``Lattice Supersymmetry,''
  Nuovo Cim.\ A {\bf 41} (1977) 1.
  %doi:10.1007/BF02730448
  %%CITATION = doi:10.1007/BF02730448;%%
  %86 citations counted in INSPIRE as of 22 Dec 2015


 
 
% 1980s

%\cite{Banks:1982ut}
\bibitem{Banks:1982ut}
  T.~Banks and P.~Windey,
  %``Supersymmetric Lattice Theories,''
  Nucl.\ Phys.\ B {\bf 198} (1982) 226.
  %doi:10.1016/0550-3213(82)90554-5
  %%CITATION = doi:10.1016/0550-3213(82)90554-5;%%
  %63 citations counted in INSPIRE as of 22 Dec 2015
  
  

%\cite{Elitzur:1982vh}
\bibitem{Elitzur:1982vh}
  S.~Elitzur, E.~Rabinovici and A.~Schwimmer,
  %``Supersymmetric Models on the Lattice,''
  Phys.\ Lett.\ B {\bf 119} (1982) 165.
  %doi:10.1016/0370-2693(82)90269-6
  %%CITATION = doi:10.1016/0370-2693(82)90269-6;%%
  %106 citations counted in INSPIRE as of 22 Dec 2015
  
%\cite{Ichinose:1982ug}
\bibitem{Ichinose:1982ug}
  I.~Ichinose,
  %``Supersymmetric Lattice Gauge Theory,''
  Phys.\ Lett.\ B {\bf 122} (1983) 68.
  %doi:10.1016/0370-2693(83)91170-X
  %%CITATION = doi:10.1016/0370-2693(83)91170-X;%%
  %29 citations counted in INSPIRE as of 22 Dec 2015
    

%\cite{Bartels:1982ue}
\bibitem{Bartels:1982ue}
  J.~Bartels and G.~Kramer,
  %``A Lattice Version of the {Wess-Zumino} Model,''
  Z.\ Phys.\ C {\bf 20} (1983) 159.
  %doi:10.1007/BF01573219
  %%CITATION = doi:10.1007/BF01573219;%%
  %35 citations counted in INSPIRE as of 22 Dec 2015



%\cite{Nojiri:1984ys}
\bibitem{Nojiri:1984ys}
  S.~Nojiri,
  %``Continuous 'Translation' and Supersymmetry on the Lattice,''
  Prog.\ Theor.\ Phys.\  {\bf 74} (1985) 819.
  %doi:10.1143/PTP.74.819
  %%CITATION = doi:10.1143/PTP.74.819;%%
  %29 citations counted in INSPIRE as of 22 Dec 2015





% WZ model and Nicolai mapping

%\cite{Sakai:1983dg}
\bibitem{Sakai:1983dg}
  N.~Sakai and M.~Sakamoto,
  %``Lattice Supersymmetry and the Nicolai Mapping,''
  Nucl.\ Phys.\ B {\bf 229} (1983) 173.
  %doi:10.1016/0550-3213(83)90359-0
  %%CITATION = doi:10.1016/0550-3213(83)90359-0;%%
  %77 citations counted in INSPIRE as of 22 Dec 2015


%\cite{Cecotti:1982ad}
\bibitem{Cecotti:1982ad}
  S.~Cecotti and L.~Girardello,
  %``Stochastic Processes in Lattice (Extended) Supersymmetry,''
  Nucl.\ Phys.\ B {\bf 226} (1983) 417.
  %doi:10.1016/0550-3213(83)90200-6
  %%CITATION = doi:10.1016/0550-3213(83)90200-6;%%
  %35 citations counted in INSPIRE as of 22 Dec 2015
  
  
%\cite{Nicolai_map}  
\bibitem{Nicolai_map}
  H.~Nicolai, Phys.\ Lett.\ B {\bf 89} (1980) 341, \ Nucl.\ Phys.\ B {\bf 176} (1980) 419.
    
%\cite{Nicolai:1979nr}
%\bibitem{Nicolai:1979nr}
%  H.~Nicolai,
  %``On a New Characterization of Scalar Supersymmetric Theories,''
%  Phys.\ Lett.\ B {\bf 89} (1980) 341.
  %doi:10.1016/0370-2693(80)90138-0
  %%CITATION = doi:10.1016/0370-2693(80)90138-0;%%
  %168 citations counted in INSPIRE as of 28 Feb 2016
  
  
%\cite{Nicolai:1980jc}
%\bibitem{Nicolai:1980jc}
%  H.~Nicolai,
  %``Supersymmetry and Functional Integration Measures,''
%  Nucl.\ Phys.\ B {\bf 176} (1980) 419.
  %doi:10.1016/0550-3213(80)90460-5
  %%CITATION = doi:10.1016/0550-3213(80)90460-5;%%
  %129 citations counted in INSPIRE as of 28 Feb 2016

%\cite{Parisi:1982ud}
\bibitem{Parisi:1982ud}
  G.~Parisi and N.~Sourlas,
  %``Supersymmetric Field Theories and Stochastic Differential Equations,''
  Nucl.\ Phys.\ B {\bf 206} (1982) 321.
  %doi:10.1016/0550-3213(82)90538-7
  %%CITATION = doi:10.1016/0550-3213(82)90538-7;%%
  %167 citations counted in INSPIRE as of 28 Feb 2016
 





%
%  SUSY QM
%


%\cite{Catterall:2000rv}
\bibitem{Catterall:2000rv}
  S.~Catterall and E.~Gregory,
  %``A Lattice path integral for supersymmetric quantum mechanics,''
  Phys.\ Lett.\ B {\bf 487} (2000) 349.
  %doi:10.1016/S0370-2693(00)00835-2
  %[hep-lat/0006013].
  %%CITATION = doi:10.1016/S0370-2693(00)00835-2;%%
  %52 citations counted in INSPIRE as of 25 Dec 2015


%\cite{Giedt:2004vb}
\bibitem{Giedt:2004vb}
  J.~Giedt, R.~Koniuk, E.~Poppitz and T.~Yavin,
  %``Less naive about supersymmetric lattice quantum mechanics,''
  JHEP {\bf 0412} (2004) 033.
  %doi:10.1088/1126-6708/2004/12/033
  %[hep-lat/0410041].
  %%CITATION = doi:10.1088/1126-6708/2004/12/033;%%
  %40 citations counted in INSPIRE as of 25 Dec 2015


%
% Relevant formulations of WZ  model
%



%\cite{Bartels:1983wm}
\bibitem{Bartels:1983wm}
  J.~Bartels and J.~B.~Bronzan,
  %``Supersymmetry On A Lattice,''
  Phys.\ Rev.\ D {\bf 28} (1983) 818.
  %doi:10.1103/PhysRevD.28.818
  %%CITATION = doi:10.1103/PhysRevD.28.818;%%
  %61 citations counted in INSPIRE as of 22 Dec 2015


%\cite{Elitzur:1983nj}
\bibitem{Elitzur:1983nj}
  S.~Elitzur and A.~Schwimmer,
  %``$N=2$ Two-dimensional {Wess-Zumino} Model on the Lattice,''
  Nucl.\ Phys.\ B {\bf 226} (1983) 109.
  %doi:10.1016/0550-3213(83)90465-0
  %%CITATION = doi:10.1016/0550-3213(83)90465-0;%%
  %62 citations counted in INSPIRE as of 22 Dec 2015
 
%\cite{Golterman:1988ta}
\bibitem{Golterman:1988ta}
  M.~F.~L.~Golterman and D.~N.~Petcher,
  %``A Local Interactive Lattice Model With Supersymmetry,''
  Nucl.\ Phys.\ B {\bf 319} (1989) 307.
  %doi:10.1016/0550-3213(89)90080-1
  %%CITATION = doi:10.1016/0550-3213(89)90080-1;%%
  %58 citations counted in INSPIRE as of 23 Dec 2015
      
  
%\cite{Bietenholz:1998qq}
\bibitem{Bietenholz:1998qq}
  W.~Bietenholz,
  %``Exact supersymmetry on the lattice,''
  Mod.\ Phys.\ Lett.\ A {\bf 14} (1999) 51.
  %doi:10.1142/S0217732399000080
  %[hep-lat/9807010].
  %%CITATION = doi:10.1142/S0217732399000080;%%
  %62 citations counted in INSPIRE as of 25 Dec 2015
    

%\cite{Fujikawa_works}
\bibitem{Fujikawa_works}
 K.~Fujikawa and M.~Ishibashi, Nucl.\ Phys.\ B {\bf 622} (2002) 115. \ K.~Fujikawa, Nucl.\ Phys.\ B {\bf 636} (2002) 80, \ Phys.\ Rev.\ D {\bf 66} (2002) 074510.
 
 
%\cite{Fujikawa:2001ka}
%\bibitem{Fujikawa:2001ka}
%  K.~Fujikawa and M.~Ishibashi,
  %``Lattice chiral symmetry and the Wess-Zumino model,''
%  Nucl.\ Phys.\ B {\bf 622} (2002) 115
  %doi:10.1016/S0550-3213(01)00592-2
  %[hep-th/0109156].
  %%CITATION = doi:10.1016/S0550-3213(01)00592-2;%%
  %37 citations counted in INSPIRE as of 25 Dec 2015



%\cite{Fujikawa:2002ic}
%\bibitem{Fujikawa:2002ic}
%  K.~Fujikawa,
  %``Supersymmetry on the lattice and the Leibniz rule,''
%  Nucl.\ Phys.\ B {\bf 636} (2002) 80
  %doi:10.1016/S0550-3213(02)00443-1
  %[hep-th/0205095].
  %%CITATION = doi:10.1016/S0550-3213(02)00443-1;%%
  %52 citations counted in INSPIRE as of 25 Dec 2015


%\cite{Catterall:2001fr}
\bibitem{Catterall:2001fr}
  S.~Catterall and S.~Karamov,
  %``Exact lattice supersymmetry: The Two-dimensional N=2 Wess-Zumino model,''
  Phys.\ Rev.\ D {\bf 65} (2002) 094501.
  %doi:10.1103/PhysRevD.65.094501
  %[hep-lat/0108024].
  %%CITATION = doi:10.1103/PhysRevD.65.094501;%%
  %90 citations counted in INSPIRE as of 25 Dec 2015


%\cite{Fujikawa:2002pa}
%\bibitem{Fujikawa:2002pa}
%  K.~Fujikawa,
  %``N = 2 Wess-Zumino model on the d = 2 Euclidean lattice,''
%  Phys.\ Rev.\ D {\bf 66} (2002) 074510
  %doi:10.1103/PhysRevD.66.074510
  %[hep-lat/0208015].
  %%CITATION = doi:10.1103/PhysRevD.66.074510;%%
  %33 citations counted in INSPIRE as of 25 Dec 2015





%\cite{Kikukawa:2002as}
\bibitem{Kikukawa:2002as}
  Y.~Kikukawa and Y.~Nakayama,
  %``Nicolai mapping versus exact chiral symmetry on the lattice,''
  Phys.\ Rev.\ D {\bf 66} (2002) 094508.
  %doi:10.1103/PhysRevD.66.094508
  %[hep-lat/0207013].
  %%CITATION = doi:10.1103/PhysRevD.66.094508;%%
  %41 citations counted in INSPIRE as of 25 Dec 2015



%\cite{Bonini:2004pm}
\bibitem{Bonini:2004pm}
  M.~Bonini and A.~Feo,
  %``Wess-Zumino model with exact supersymmetry on the lattice,''
  JHEP {\bf 0409} (2004) 011.
  %doi:10.1088/1126-6708/2004/09/011
  %[hep-lat/0402034].
  %%CITATION = doi:10.1088/1126-6708/2004/09/011;%%
  %34 citations counted in INSPIRE as of 25 Dec 2015


%\cite{Giedt:2004qs}
\bibitem{Giedt:2004qs}
  J.~Giedt and E.~Poppitz,
  %``Lattice supersymmetry, superfields and renormalization,''
  JHEP {\bf 0409} (2004) 029.
  %doi:10.1088/1126-6708/2004/09/029
  %[hep-th/0407135].
  %%CITATION = doi:10.1088/1126-6708/2004/09/029;%%
  %39 citations counted in INSPIRE as of 25 Dec 2015




%\cite{Kikukawa:2004dd}
\bibitem{Kikukawa:2004dd}
  Y.~Kikukawa and H.~Suzuki,
  %``A Local formulation of lattice Wess-Zumino model with exact U(1)(R) symmetry,''
  JHEP {\bf 0502} (2005) 012.
  %doi:10.1088/1126-6708/2005/02/012
  %[hep-lat/0412042].
  %%CITATION = doi:10.1088/1126-6708/2005/02/012;%%
  %20 citations counted in INSPIRE as of 25 Dec 2015


%\cite{Kadoh:2010ca}
\bibitem{Kadoh:2010ca}
  D.~Kadoh and H.~Suzuki,
  %``Supersymmetry restoration in lattice formulations of 2D $\mathcal{N}=(2,2)$ WZ model based on the Nicolai map,''
  Phys.\ Lett.\ B {\bf 696} (2011) 163.
  %doi:10.1016/j.physletb.2010.12.012
  %[arXiv:1011.0788 [hep-lat]].
  %%CITATION = doi:10.1016/j.physletb.2010.12.012;%%
  %2 citations counted in INSPIRE as of 25 Dec 2015
        




% Numerical simualations of WZ model


%\cite{Beccaria:1998vi}
\bibitem{Beccaria:1998vi}
  M.~Beccaria, G.~Curci and E.~D'Ambrosio,
  %``Simulation of supersymmetric models with a local Nicolai map,''
  Phys.\ Rev.\ D {\bf 58} (1998) 065009.
  %doi:10.1103/PhysRevD.58.065009
  %[hep-lat/9804010].
  %%CITATION = doi:10.1103/PhysRevD.58.065009;%%
  %24 citations counted in INSPIRE as of 25 Dec 2015


%\cite{Catterall:2003ae}
\bibitem{Catterall:2003ae}
  S.~Catterall and S.~Karamov,
  %``A Lattice study of the two-dimensional Wess-Zumino model,''
  Phys.\ Rev.\ D {\bf 68} (2003) 014503.
  %doi:10.1103/PhysRevD.68.014503
  %[hep-lat/0305002].
  %%CITATION = doi:10.1103/PhysRevD.68.014503;%%
  %55 citations counted in INSPIRE as of 25 Dec 2015




%\cite{Bergner:2007pu}
\bibitem{Bergner:2007pu}
  G.~Bergner, T.~Kaestner, S.~Uhlmann and A.~Wipf,
  %``Low-dimensional Supersymmetric Lattice Models,''
  Annals Phys.\  {\bf 323} (2008) 946.
  %doi:10.1016/j.aop.2007.06.010
  %[arXiv:0705.2212 [hep-lat]].
  %%CITATION = doi:10.1016/j.aop.2007.06.010;%%
  %37 citations counted in INSPIRE as of 25 Dec 2015

  

%\cite{Kastner:2008zc}
\bibitem{Kastner:2008zc}
  T.~Kastner, G.~Bergner, S.~Uhlmann, A.~Wipf and C.~Wozar,
  %``Two-Dimensional Wess-Zumino Models at Intermediate Couplings,''
  Phys.\ Rev.\ D {\bf 78} (2008) 095001.
  %doi:10.1103/PhysRevD.78.095001
  %[arXiv:0807.1905 [hep-lat]].
  %%CITATION = doi:10.1103/PhysRevD.78.095001;%%
  %21 citations counted in INSPIRE as of 25 Dec 2015
  
  

%\cite{Yu:2010zv}
\bibitem{Yu:2010zv}
  Y.~Yu and K.~Yang,
  %``Simulating Wess-Zumino Supersymmetry Model in Optical Lattices,''
  Phys.\ Rev.\ Lett.\  {\bf 105} (2010) 150605.
  %doi:10.1103/PhysRevLett.105.150605
  %[arXiv:1005.1399 [cond-mat.quant-gas]].
  %%CITATION = doi:10.1103/PhysRevLett.105.150605;%%
  %19 citations counted in INSPIRE as of 25 Dec 2015  
   





% Sigma model


%\cite{Catterall_:Ghadab}
\bibitem{Catterall_:Ghadab}
  S.~Catterall and S.~Ghadab,  JHEP {\bf 0405} (2004) 044, \ JHEP {\bf 0610} (2006) 063.

%\cite{Catterall:2003uf}
%\bibitem{Catterall:2003uf}
%  S.~Catterall and S.~Ghadab,
  %``Lattice sigma models with exact supersymmetry,''
%  JHEP {\bf 0405} (2004) 044.
  %doi:10.1088/1126-6708/2004/05/044
  %[hep-lat/0311042].
  %%CITATION = doi:10.1088/1126-6708/2004/05/044;%%
  %59 citations counted in INSPIRE as of 25 Dec 2015


%\cite{Catterall:2006sj}
%\bibitem{Catterall:2006sj}
%  S.~Catterall and S.~Ghadab,
  %``Twisted supersymmetric sigma model on the lattice,''
%  JHEP {\bf 0610} (2006) 063
  %doi:10.1088/1126-6708/2006/10/063
  %[hep-lat/0607010].
  %%CITATION = doi:10.1088/1126-6708/2006/10/063;%%
  %21 citations counted in INSPIRE as of 25 Dec 2015
  

%\cite{Flore:2012xj}
\bibitem{Flore:2012xj}
  R.~Flore, D.~Korner, A.~Wipf and C.~Wozar,
  %``Supersymmetric Nonlinear O(3) Sigma Model on the Lattice,''
  JHEP {\bf 1211} (2012) 159.
  %doi:10.1007/JHEP11(2012)159
  %[arXiv:1207.6947 [hep-lat]].
  %%CITATION = doi:10.1007/JHEP11(2012)159;%%
  %3 citations counted in INSPIRE as of 25 Dec 2015

 
   
   
     
% DESY Munster
  

%\cite{Montvay_works}
\bibitem{Montvay_works}
  I.~Montvay, Nucl.\ Phys.\ B {\bf 466} (1996) 259, \ Int.\ J.\ Mod.\ Phys.\ A {\bf 17} (2002) 2377.

%\cite{Montvay:1995ea}
%\bibitem{Montvay:1995ea}
%  I.~Montvay,
  %``An algorithm for gluinos on the lattice,''
%  Nucl.\ Phys.\ B {\bf 466} (1996) 259
  %doi:10.1016/0550-3213(96)00086-7
  %[hep-lat/9510042].
  %%CITATION = doi:10.1016/0550-3213(96)00086-7;%%
  %126 citations counted in INSPIRE as of 25 Dec 2015


%\cite{Montvay:2001aj}
%\bibitem{Montvay:2001aj}
%  I.~Montvay,
  %``Supersymmetric Yang-Mills theory on the lattice,''
%  Int.\ J.\ Mod.\ Phys.\ A {\bf 17} (2002) 2377
  %doi:10.1142/S0217751X0201090X
  %[hep-lat/0112007].
  %%CITATION = doi:10.1142/S0217751X0201090X;%%
  %79 citations counted in INSPIRE as of 25 Dec 2015



%\cite{Curci:1986sm}
\bibitem{Curci:1986sm}
  G.~Curci and G.~Veneziano,
  %``Supersymmetry and the Lattice: A Reconciliation?,''
  Nucl.\ Phys.\ B {\bf 292} (1987) 555.
  %doi:10.1016/0550-3213(87)90660-2
  %%CITATION = doi:10.1016/0550-3213(87)90660-2;%%
  %150 citations counted in INSPIRE as of 23 Dec 2015
  


%\cite{Suzuki:2012pc}
\bibitem{Suzuki:2012pc}
  H.~Suzuki,
  %``Supersymmetry, chiral symmetry and the generalized BRS transformation in lattice formulations of 4D $\mathcal{N}=1$ SYM,''
  Nucl.\ Phys.\ B {\bf 861} (2012) 290.
  %doi:10.1016/j.nuclphysb.2012.04.008
  %[arXiv:1202.2598 [hep-lat]].
  %%CITATION = doi:10.1016/j.nuclphysb.2012.04.008;%%
  %11 citations counted in INSPIRE as of 25 Dec 2015



%\cite{desy_munster_hp}
 \bibitem{desy_munster_hp}
 The DESY-Munster collaboration :   
 http://pauli.uni-muenster.de/~munsteg/susy.html   
 
%\cite{desy_munster_recent}
\bibitem{desy_munster_recent}
  G.~Bergner, P.~Giudice, I.~Montvay, G.~Munster and S.~Piemonte, PoS LATTICE {\bf 2015} (2015) 240, \ arXiv:1512.07014 [hep-lat].



%\cite{Bergner:2015lba}
%\bibitem{Bergner:2015lba}
%  G.~Bergner, P.~Giudice, I.~Montvay, G.~Munster and S.~Piemonte,
  %``Supermultiplets of the N=1 supersymmetric Yang-Mills theory in the continuum limit,''
%  PoS LATTICE {\bf 2015} (2015) 240
  %[arXiv:1510.08795 [hep-lat]].
  %%CITATION = ARXIV:1510.08795;%%
  %1 citations counted in INSPIRE as of 28 Feb 2016
 
 
%\cite{Bergner:2015adz}
%\bibitem{Bergner:2015adz}
%  G.~Bergner, P.~Giudice, I.~Montvay, G.~Munster and S.~Piemonte,
  %``The light bound states of supersymmetric SU(2) Yang-Mills theory,''
%  arXiv:1512.07014 [hep-lat].
  %%CITATION = ARXIV:1512.07014;%%

            
%\cite{Veneziano:1982ah}
\bibitem{Veneziano:1982ah}
  G.~Veneziano and S.~Yankielowicz,
  %``An Effective Lagrangian for the Pure N=1 Supersymmetric Yang-Mills Theory,''
  Phys.\ Lett.\ B {\bf 113} (1982) 231.
  %doi:10.1016/0370-2693(82)90828-0
  %%CITATION = doi:10.1016/0370-2693(82)90828-0;%%
  %667 citations counted in INSPIRE as of 20 janv. 2016
  
  


% The other group of N=1 SYM with Wilson fermion 
  
%\cite{Donini:1997hh}
%\bibitem{Donini:1997hh}
%  A.~Donini, M.~Guagnelli, P.~Hernandez and A.~Vladikas,
  %``Towards N=1 superYang-Mills on the lattice,''
%  Nucl.\ Phys.\ B {\bf 523} (1998) 529
  %doi:10.1016/S0550-3213(98)00166-7
  %[hep-lat/9710065].
  %%CITATION = doi:10.1016/S0550-3213(98)00166-7;%%
  %56 citations counted in INSPIRE as of 25 Dec 2015






% Ginsparg-Wison fermions

%\cite{Kaplan:1992bt}
\bibitem{Kaplan:1992bt}
  D.~B.~Kaplan,
  %``A Method for simulating chiral fermions on the lattice,''
  Phys.\ Lett.\ B {\bf 288} (1992) 342.
  %doi:10.1016/0370-2693(92)91112-M
  %[hep-lat/9206013].
  %%CITATION = doi:10.1016/0370-2693(92)91112-M;%%
  %981 citations counted in INSPIRE as of 01 Mar 2016


%\cite{Shamir:1993zy}
\bibitem{Shamir:1993zy}
  Y.~Shamir,
  %``Chiral fermions from lattice boundaries,''
  Nucl.\ Phys.\ B {\bf 406} (1993) 90.
  %doi:10.1016/0550-3213(93)90162-I
  %[hep-lat/9303005].
  %%CITATION = doi:10.1016/0550-3213(93)90162-I;%%
  %588 citations counted in INSPIRE as of 01 Mar 2016


%\cite{Furman:1994ky}
\bibitem{Furman:1994ky}
  V.~Furman and Y.~Shamir,
  %``Axial symmetries in lattice QCD with Kaplan fermions,''
  Nucl.\ Phys.\ B {\bf 439} (1995) 54.
  %doi:10.1016/0550-3213(95)00031-M
  %[hep-lat/9405004].
  %%CITATION = doi:10.1016/0550-3213(95)00031-M;%%
  %518 citations counted in INSPIRE as of 01 Mar 2016



%\cite{Neuberger_works}
\bibitem{Neuberger_works}
  H.~Neuberger,  \ Phys.\ Lett.\ B {\bf 417} (1998) 141, \ Phys.\ Lett.\ B {\bf 427} (1998) 353.

%\cite{Neuberger:1997fp}
%\bibitem{Neuberger:1997fp}
%  H.~Neuberger,
  %``Exactly massless quarks on the lattice,''
%  Phys.\ Lett.\ B {\bf 417} (1998) 141
  %doi:10.1016/S0370-2693(97)01368-3
  %[hep-lat/9707022].
  %%CITATION = doi:10.1016/S0370-2693(97)01368-3;%%
  %1214 citations counted in INSPIRE as of 20 janv. 2016


%\cite{Neuberger:1998wv}
%\bibitem{Neuberger:1998wv}
%  H.~Neuberger,
  %``More about exactly massless quarks on the lattice,''
%  Phys.\ Lett.\ B {\bf 427} (1998) 353
  %doi:10.1016/S0370-2693(98)00355-4
  %[hep-lat/9801031].
  %%CITATION = doi:10.1016/S0370-2693(98)00355-4;%%
  %681 citations counted in INSPIRE as of 20 janv. 2016



%\cite{Ginsparg:1981bj}
\bibitem{Ginsparg:1981bj}
  P.~H.~Ginsparg and K.~G.~Wilson,
  %``A Remnant of Chiral Symmetry on the Lattice,''
  Phys.\ Rev.\ D {\bf 25} (1982) 2649.
  %doi:10.1103/PhysRevD.25.2649
  %%CITATION = doi:10.1103/PhysRevD.25.2649;%%
  %924 citations counted in INSPIRE as of 20 janv. 2016


%\cite{Luscher:1998pqa}
\bibitem{Luscher:1998pqa}
  M.~Luscher,
  %``Exact chiral symmetry on the lattice and the Ginsparg-Wilson relation,''
  Phys.\ Lett.\ B {\bf 428} (1998) 342.
  %doi:10.1016/S0370-2693(98)00423-7
  %[hep-lat/9802011].
  %%CITATION = doi:10.1016/S0370-2693(98)00423-7;%%
  %735 citations counted in INSPIRE as of 01 Mar 2016


%\cite{Nielsen_Ninomiya}
\bibitem{Nielsen_Ninomiya}
  H.~B.~Nielsen and M.~Ninomiya, \ Nucl.\ Phys.\ B {\bf 185} (1981) 20
   [Nucl.\ Phys.\ B {\bf 195} (1982) 541], \ Nucl.\ Phys.\ B {\bf 193} (1981) 173.


%\cite{Nielsen:1980rz}
%\bibitem{Nielsen:1980rz}
%  H.~B.~Nielsen and M.~Ninomiya,
  %``Absence of Neutrinos on a Lattice. 1. Proof by Homotopy Theory,''
%  Nucl.\ Phys.\ B {\bf 185} (1981) 20
%   [Nucl.\ Phys.\ B {\bf 195} (1982) 541].
  %doi:10.1016/0550-3213(81)90361-8
  %%CITATION = doi:10.1016/0550-3213(81)90361-8;%%
  %926 citations counted in INSPIRE as of 20 janv. 2016


%\cite{Nielsen:1981xu}
%\bibitem{Nielsen:1981xu}
%  H.~B.~Nielsen and M.~Ninomiya,
  %``Absence of Neutrinos on a Lattice. 2. Intuitive Topological Proof,''
%  Nucl.\ Phys.\ B {\bf 193} (1981) 173.
  %doi:10.1016/0550-3213(81)90524-1
  %%CITATION = doi:10.1016/0550-3213(81)90524-1;%%
  %596 citations counted in INSPIRE as of 20 janv. 2016


%\cite{Nishimura_works}
\bibitem{Nishimura_works}
  J.~Nishimura, \ Phys.\ Lett.\ B {\bf 406} (1997) 215, \ N.~Maru and J.~Nishimura, \  Int.\ J.\ Mod.\ Phys.\ A {\bf 13} (1998) 2841.


%\cite{Nishimura:1997vg}
%\bibitem{Nishimura:1997vg}
%  J.~Nishimura,
  %``Four-dimensional N=1 supersymmetric Yang-Mills theory on the lattice without fine tuning,''
%  Phys.\ Lett.\ B {\bf 406} (1997) 215
  %doi:10.1016/S0370-2693(97)00674-6
  %[hep-lat/9701013].
  %%CITATION = doi:10.1016/S0370-2693(97)00674-6;%%
  %58 citations counted in INSPIRE as of 25 Dec 2015


%\cite{Maru:1997kh}
%\bibitem{Maru:1997kh}
%  N.~Maru and J.~Nishimura,
  %``Lattice formulation of supersymmetric Yang-Mills theories without fine tuning,''
%  Int.\ J.\ Mod.\ Phys.\ A {\bf 13} (1998) 2841
  %doi:10.1142/S0217751X9800144X
  %[hep-th/9705152].
  %%CITATION = doi:10.1142/S0217751X9800144X;%%
  %56 citations counted in INSPIRE as of 25 Dec 2015
  
  

%\cite{Neuberger:1997bg}
\bibitem{Neuberger:1997bg}
  H.~Neuberger,
  %``Vector - like gauge theories with almost massless fermions on the lattice,''
  Phys.\ Rev.\ D {\bf 57} (1998) 5417.
  %doi:10.1103/PhysRevD.57.5417
  %[hep-lat/9710089].
  %%CITATION = doi:10.1103/PhysRevD.57.5417;%%
  %295 citations counted in INSPIRE as of 25 Dec 2015
  
  
  
  
 %\cite{Aoyama:1998in}
\bibitem{Aoyama:1998in}
  T.~Aoyama and Y.~Kikukawa,
  %``Overlap formula for the chiral multiplet,''
  Phys.\ Rev.\ D {\bf 59} (1999) 054507.
  %doi:10.1103/PhysRevD.59.054507
  %[hep-lat/9803016].
  %%CITATION = doi:10.1103/PhysRevD.59.054507;%%
  %25 citations counted in INSPIRE as of 25 Dec 2015
  



%\cite{Kaplan:1999jn}
\bibitem{Kaplan:1999jn}
  D.~B.~Kaplan and M.~Schmaltz,
  %``Supersymmetric Yang-Mills theories from domain wall fermions,''
  Chin.\ J.\ Phys.\  {\bf 38} (2000) 543.
  %[hep-lat/0002030].
  %%CITATION = HEP-LAT/0002030;%%
  %69 citations counted in INSPIRE as of 25 Dec 2015
  
  

%\cite{Fleming:2000fa}
\bibitem{Fleming:2000fa}
  G.~T.~Fleming, J.~B.~Kogut and P.~M.~Vranas,
  %``SuperYang-Mills on the lattice with domain wall fermions,''
  Phys.\ Rev.\ D {\bf 64} (2001) 034510.
  %doi:10.1103/PhysRevD.64.034510
  %[hep-lat/0008009].
  %%CITATION = doi:10.1103/PhysRevD.64.034510;%%
  %83 citations counted in INSPIRE as of 25 Dec 2015



%\cite{Giedt:2008xm}
\bibitem{Giedt:2008xm}
  J.~Giedt, R.~Brower, S.~Catterall, G.~T.~Fleming and P.~Vranas,
  %``Lattice super-Yang-Mills using domain wall fermions in the chiral limit,''
  Phys.\ Rev.\ D {\bf 79} (2009) 025015.
  %doi:10.1103/PhysRevD.79.025015
  %[arXiv:0810.5746 [hep-lat]].
  %%CITATION = doi:10.1103/PhysRevD.79.025015;%%
  %47 citations counted in INSPIRE as of 25 Dec 2015



%\cite{Endres:2009yp}
\bibitem{Endres:2009yp}
  M.~G.~Endres,
  %``Dynamical simulation of N=1 supersymmetric Yang-Mills theory with domain wall fermions,''
  Phys.\ Rev.\ D {\bf 79} (2009) 094503.
  %doi:10.1103/PhysRevD.79.094503
  %[arXiv:0902.4267 [hep-lat]].
  %%CITATION = doi:10.1103/PhysRevD.79.094503;%%
  %37 citations counted in INSPIRE as of 25 Dec 2015




% CKKU method


%\cite{Cohen:2003xe}
\bibitem{Cohen:2003xe}
  A.~G.~Cohen, D.~B.~Kaplan, E.~Katz and M.~Unsal,
  %``Supersymmetry on a Euclidean space-time lattice. 1. A Target theory with four supercharges,''
  JHEP {\bf 0308} (2003) 024.
  %doi:10.1088/1126-6708/2003/08/024
  %[hep-lat/0302017].
  %%CITATION = doi:10.1088/1126-6708/2003/08/024;%%
  %146 citations counted in INSPIRE as of 25 Dec 2015
  

  
%\cite{Cohen:2003qw}
\bibitem{Cohen:2003qw}
  A.~G.~Cohen, D.~B.~Kaplan, E.~Katz and M.~Unsal,
  %``Supersymmetry on a Euclidean space-time lattice. 2. Target theories with eight supercharges,''
  JHEP {\bf 0312} (2003) 031.
  %doi:10.1088/1126-6708/2003/12/031
  %[hep-lat/0307012].
  %%CITATION = doi:10.1088/1126-6708/2003/12/031;%%
  %121 citations counted in INSPIRE as of 25 Dec 2015  
  


%\cite{Kaplan:2005ta}
\bibitem{Kaplan:2005ta}
  D.~B.~Kaplan and M.~Unsal,
  %``A Euclidean lattice construction of supersymmetric Yang-Mills theories with sixteen supercharges,''
  JHEP {\bf 0509} (2005) 042.
  %doi:10.1088/1126-6708/2005/09/042
  %[hep-lat/0503039].
  %%CITATION = doi:10.1088/1126-6708/2005/09/042;%%
  %121 citations counted in INSPIRE as of 25 Dec 2015



%\cite{Unsal:2008kx}
\bibitem{Unsal:2008kx}
  M.~Unsal,
  %``Deformed matrix models, supersymmetric lattice twists and N = 1/4 supersymmetry,''
  JHEP {\bf 0905} (2009) 082.
  %doi:10.1088/1126-6708/2009/05/082
  %[arXiv:0809.3216 [hep-lat]].
  %%CITATION = doi:10.1088/1126-6708/2009/05/082;%%
  %8 citations counted in INSPIRE as of 25 Dec 2015
  
  


%\cite{Unsal:2005yh}
\bibitem{Unsal:2005yh}
  M.~Unsal,
  %``Compact gauge fields for supersymmetric lattices,''
  JHEP {\bf 0511} (2005) 013.
  %doi:10.1088/1126-6708/2005/11/013
  %[hep-lat/0504016].
  %%CITATION = doi:10.1088/1126-6708/2005/11/013;%%
  %29 citations counted in INSPIRE as of 25 Dec 2015


%\cite{Damgaard:2007be}
\bibitem{Damgaard:2007be}
  P.~H.~Damgaard and S.~Matsuura,
  %``Classification of supersymmetric lattice gauge theories by orbifolding,''
  JHEP {\bf 0707} (2007) 051.
  %doi:10.1088/1126-6708/2007/07/051
  %[arXiv:0704.2696 [hep-lat]].
  %%CITATION = doi:10.1088/1126-6708/2007/07/051;%%
  %38 citations counted in INSPIRE as of 25 Dec 2015


%\cite{Endres:2006ic}
\bibitem{Endres:2006ic}
  M.~G.~Endres and D.~B.~Kaplan,
  %``Lattice formulation of (2,2) supersymmetric gauge theories with matter fields,''
  JHEP {\bf 0610} (2006) 076.
  %doi:10.1088/1126-6708/2006/10/076
  %[hep-lat/0604012].
  %%CITATION = doi:10.1088/1126-6708/2006/10/076;%%
  %42 citations counted in INSPIRE as of 25 Dec 2015


 
%\cite{Matsuura:2008cfa}
\bibitem{Matsuura:2008cfa}
  S.~Matsuura,
  %``Two-dimensional N=(2,2) Supersymmetric Lattice Gauge Theory with Matter Fields in the Fundamental Representation,''
  JHEP {\bf 0807} (2008) 127.
  %doi:10.1088/1126-6708/2008/07/127
  %[arXiv:0805.4491 [hep-th]].
  %%CITATION = doi:10.1088/1126-6708/2008/07/127;%%
  %22 citations counted in INSPIRE as of 25 Dec 2015



%\cite{Joseph_works}
\bibitem{Joseph_works}
A.~Joseph, \ JHEP {\bf 1309} (2013) 046, \ JHEP {\bf 1401} (2014) 093, \ JHEP {\bf 1407} (2014) 067.



%\cite{Joseph:2013jya}
%\bibitem{Joseph:2013jya}
%  A.~Joseph,
  %``Lattice formulation of three-dimensional ${\cal N}=4$ gauge theory with fundamental matter fields,''
%  JHEP {\bf 1309} (2013) 046
  %doi:10.1007/JHEP09(2013)046
  %[arXiv:1307.3281 [hep-lat]].
  %%CITATION = doi:10.1007/JHEP09(2013)046;%%
  %9 citations counted in INSPIRE as of 25 Dec 2015


%\cite{Joseph:2013bra}
%\bibitem{Joseph:2013bra}
%  A.~Joseph,
  %``Supersymmetric quiver gauge theories on the lattice,''
%  JHEP {\bf 1401} (2014) 093
  %doi:10.1007/JHEP01(2014)093
  %[arXiv:1311.5111 [hep-lat]].
  %%CITATION = doi:10.1007/JHEP01(2014)093;%%
  %5 citations counted in INSPIRE as of 25 Dec 2015
  
  
%\cite{Joseph:2014bwa}
%\bibitem{Joseph:2014bwa}
%  A.~Joseph,
  %``Two-dimensional $ \mathcal{N} $ = (2, 2) lattice gauge theories with matter in higher representations,''
%  JHEP {\bf 1407} (2014) 067
  %doi:10.1007/JHEP07(2014)067
  %[arXiv:1403.4390 [hep-lat]].
  %%CITATION = doi:10.1007/JHEP07(2014)067;%%
  %3 citations counted in INSPIRE as of 25 Dec 2015



%\cite{Giedt_works}
\bibitem{Giedt_works}
  J.~Giedt, \ Nucl.\ Phys.\ B {\bf 668} (2003) 138, \ Nucl.\ Phys.\ B {\bf 674} (2003) 259, \ Int.\ J.\ Mod.\ Phys.\ A {\bf 21} (2006) 3039.


%\cite{Giedt:2003ve}
%\bibitem{Giedt:2003ve}
%  J.~Giedt,
  %``Nonpositive fermion determinants in lattice supersymmetry,''
%  Nucl.\ Phys.\ B {\bf 668} (2003) 138
  %doi:10.1016/j.nuclphysb.2003.07.006
  %[hep-lat/0304006].
  %%CITATION = doi:10.1016/j.nuclphysb.2003.07.006;%%
  %55 citations counted in INSPIRE as of 25 Dec 2015


%\cite{Giedt:2003vy}
%\bibitem{Giedt:2003vy}
%  J.~Giedt,
  %``The Fermion determinant in (4,4) 2-d lattice superYang-Mills,''
%  Nucl.\ Phys.\ B {\bf 674} (2003) 259
  %doi:10.1016/j.nuclphysb.2003.09.045
  %[hep-lat/0307024].
  %%CITATION = doi:10.1016/j.nuclphysb.2003.09.045;%%
  %45 citations counted in INSPIRE as of 25 Dec 2015
  



%\cite{Giedt:2006pd}
%\bibitem{Giedt:2006pd}
%  J.~Giedt,
  %``Deconstruction and other approaches to supersymmetric lattice field theories,''
%  Int.\ J.\ Mod.\ Phys.\ A {\bf 21} (2006) 3039
  %doi:10.1142/S0217751X06031752
  %[hep-lat/0602007].
  %%CITATION = doi:10.1142/S0217751X06031752;%%
  %57 citations counted in INSPIRE as of 25 Dec 2015



%\cite{Onogi:2005cz}
%\bibitem{Onogi:2005cz}
%  T.~Onogi and T.~Takimi,
  %``Perturbative study of the supersymmetric lattice theory from matrix model,''
%  Phys.\ Rev.\ D {\bf 72} (2005) 074504
  %doi:10.1103/PhysRevD.72.074504
  %[hep-lat/0506014].
  %%CITATION = doi:10.1103/PhysRevD.72.074504;%%
  %30 citations counted in INSPIRE as of 25 Dec 2015
  
    
%\cite{Catterall:2008dv}
\bibitem{Catterall:2008dv}
  S.~Catterall,
  %``First results from simulations of supersymmetric lattices,''
  JHEP {\bf 0901} (2009) 040.
  %doi:10.1088/1126-6708/2009/01/040
  %[arXiv:0811.1203 [hep-lat]].
  %%CITATION = doi:10.1088/1126-6708/2009/01/040;%%
  %28 citations counted in INSPIRE as of 25 Dec 2015
  
   
  

% Sugino method



%\cite{Sugino:2003yb}
\bibitem{Sugino:2003yb}
  F.~Sugino,
  %``A Lattice formulation of superYang-Mills theories with exact supersymmetry,''
  JHEP {\bf 0401} (2004) 015.
  %doi:10.1088/1126-6708/2004/01/015
  %[hep-lat/0311021].
  %%CITATION = doi:10.1088/1126-6708/2004/01/015;%%
  %124 citations counted in INSPIRE as of 25 Dec 2015


%\cite{Sugino:2004qd}
\bibitem{Sugino:2004qd}
  F.~Sugino,
  %``SuperYang-Mills theories on the two-dimensional lattice with exact supersymmetry,''
  JHEP {\bf 0403} (2004) 067.
  %doi:10.1088/1126-6708/2004/03/067
  %[hep-lat/0401017].
  %%CITATION = doi:10.1088/1126-6708/2004/03/067;%%
  %114 citations counted in INSPIRE as of 25 Dec 2015


%\cite{Sugino:2004uv}
\bibitem{Sugino:2004uv}
  F.~Sugino,
  %``Various super Yang-Mills theories with exact supersymmetry on the lattice,''
  JHEP {\bf 0501} (2005) 016.
  %doi:10.1088/1126-6708/2005/01/016
  %[hep-lat/0410035].
  %%CITATION = doi:10.1088/1126-6708/2005/01/016;%%
  %79 citations counted in INSPIRE as of 25 Dec 2015


%\cite{Sugino:2006uf}
\bibitem{Sugino:2006uf}
  F.~Sugino,
  %``Two-dimensional compact N=(2,2) lattice super Yang-Mills theory with exact supersymmetry,''
  Phys.\ Lett.\ B {\bf 635} (2006) 218.
  %doi:10.1016/j.physletb.2006.02.064
  %[hep-lat/0601024].
  %%CITATION = doi:10.1016/j.physletb.2006.02.064;%%
  %54 citations counted in INSPIRE as of 25 Dec 2015




%\cite{TFT}    
\bibitem{TFT}
  R.~Dijkgraaf and G.~W.~Moore, Commun.\ Math.\ Phys.\  {\bf 185} (1997) 411, \ 
  C.~Vafa and E.~Witten,  Nucl.\ Phys.\ B {\bf 431} (1994) 3, \ 
  J.~M.~F.~Labastida and C.~Lozano,  Nucl.\ Phys.\ B {\bf 502} (1997) 741, \ 
  M.~Blau and G.~Thompson, Nucl.\ Phys.\ B {\bf 492} (1997) 545.
  
  

%\cite{Sugino:2008yp}
\bibitem{Sugino:2008yp}
  F.~Sugino,
  %``Lattice Formulation of Two-Dimensional N=(2,2) SQCD with Exact Supersymmetry,''
  Nucl.\ Phys.\ B {\bf 808} (2009) 292.
  %doi:10.1016/j.nuclphysb.2008.09.035
  %[arXiv:0807.2683 [hep-lat]].
  %%CITATION = doi:10.1016/j.nuclphysb.2008.09.035;%%
  %23 citations counted in INSPIRE as of 25 Dec 2015



%\cite{Kikukawa:2008xw}
\bibitem{Kikukawa:2008xw}
  Y.~Kikukawa and F.~Sugino,
  %``Ginsparg-Wilson Formulation of 2D N = (2,2) SQCD with Exact Lattice Supersymmetry,''
  Nucl.\ Phys.\ B {\bf 819} (2009) 76.
  %doi:10.1016/j.nuclphysb.2009.04.007
  %[arXiv:0811.0916 [hep-lat]].
  %%CITATION = doi:10.1016/j.nuclphysb.2009.04.007;%%
  %18 citations counted in INSPIRE as of 25 Dec 2015



%\cite{Kadoh:2009yf}
\bibitem{Kadoh:2009yf}
  D.~Kadoh, F.~Sugino and H.~Suzuki,
  %``Lattice formulation of 2D N = (2,2) SQCD based on the B model twist,''
  Nucl.\ Phys.\ B {\bf 820} (2009) 99.
  %doi:10.1016/j.nuclphysb.2009.05.012
  %[arXiv:0903.5398 [hep-lat]].
  %%CITATION = doi:10.1016/j.nuclphysb.2009.05.012;%%
  %15 citations counted in INSPIRE as of 25 Dec 2015
  
  

%\cite{Matsuura:2014pua}
\bibitem{Matsuura:2014pua}
  S.~Matsuura and F.~Sugino,
  %``Lattice formulation for 2d = (2, 2), (4, 4) super Yang-Mills theories without admissibility conditions,''
  JHEP {\bf 1404} (2014) 088.
  %doi:10.1007/JHEP04(2014)088
  %[arXiv:1402.0952 [hep-lat]].
  %%CITATION = doi:10.1007/JHEP04(2014)088;%%
  %5 citations counted in INSPIRE as of 25 Dec 2015









% Numerical studies usign Sugino models

%\cite{Kanamori_Suzuki}
\bibitem{Kanamori_Suzuki}
H.~Suzuki, JHEP {\bf 0709} (2007) 052, \  I.~Kanamori and H.~Suzuki, Nucl.\ Phys.\ B {\bf 811} (2009) 420, \ Phys.\ Lett.\ B {\bf 672} (2009) 307.
 
 
%\cite{Suzuki:2007jt}
%\bibitem{Suzuki:2007jt}
%  H.~Suzuki,
  %``Two-dimensional N = (2,2) super Yang-Mills theory on computer,''
%  JHEP {\bf 0709} (2007) 052
  %doi:10.1088/1126-6708/2007/09/052
  %[arXiv:0706.1392 [hep-lat]].
  %%CITATION = doi:10.1088/1126-6708/2007/09/052;%%
  %26 citations counted in INSPIRE as of 25 Dec 2015

  
  
%\cite{Kanamori:2008bk}
%\bibitem{Kanamori:2008bk}
%  I.~Kanamori and H.~Suzuki,
  %``Restoration of supersymmetry on the lattice: Two-dimensional N = (2,2) supersymmetric Yang-Mills theory,''
%  Nucl.\ Phys.\ B {\bf 811} (2009) 420
  %doi:10.1016/j.nuclphysb.2008.11.021
  %[arXiv:0809.2856 [hep-lat]].
  %%CITATION = doi:10.1016/j.nuclphysb.2008.11.021;%%
  %48 citations counted in INSPIRE as of 25 Dec 2015



%\cite{Kanamori:2008yy}
%\bibitem{Kanamori:2008yy}
%  I.~Kanamori and H.~Suzuki,
  %``Some physics of the two-dimensional N = (2,2) supersymmetric Yang-Mills theory: Lattice Monte Carlo study,''
%  Phys.\ Lett.\ B {\bf 672} (2009) 307
  %doi:10.1016/j.physletb.2009.01.039
  %[arXiv:0811.2851 [hep-lat]].
  %%CITATION = doi:10.1016/j.physletb.2009.01.039;%%
  %13 citations counted in INSPIRE as of 25 Dec 2015
    

     

%\cite{Hanada:2009hq}
\bibitem{Hanada:2009hq}
  M.~Hanada and I.~Kanamori,
  %``Lattice study of two-dimensional N=(2,2) super Yang-Mills at large-N,''
  Phys.\ Rev.\ D {\bf 80} (2009) 065014.
  %doi:10.1103/PhysRevD.80.065014
  %[arXiv:0907.4966 [hep-lat]].
  %%CITATION = doi:10.1103/PhysRevD.80.065014;%%
  %37 citations counted in INSPIRE as of 25 Dec 2015     
  
  



% link approach

%\cite{link_approach}
\bibitem{link_approach}
  A.~D'Adda, I.~Kanamori, N.~Kawamoto and K.~Nagata, \ Nucl.\ Phys.\ B {\bf 707} (2005) 100, \ Phys.\ Lett.\ B {\bf 633} (2006) 645, \ Nucl.\ Phys.\ B {\bf 798} (2008) 168.

%\cite{D'Adda:2004jb}
%\bibitem{D'Adda:2004jb}
%  A.~D'Adda, I.~Kanamori, N.~Kawamoto and K.~Nagata,
  %``Twisted superspace on a lattice,''
%  Nucl.\ Phys.\ B {\bf 707} (2005) 100
  %doi:10.1016/j.nuclphysb.2004.11.046
  %[hep-lat/0406029].
  %%CITATION = doi:10.1016/j.nuclphysb.2004.11.046;%%
  %77 citations counted in INSPIRE as of 25 Dec 2015


%\cite{D'Adda:2005zk}
%\bibitem{D'Adda:2005zk}
%  A.~D'Adda, I.~Kanamori, N.~Kawamoto and K.~Nagata,
  %``Exact extended supersymmetry on a lattice: Twisted N=2 super Yang-Mills in two dimensions,''
%  Phys.\ Lett.\ B {\bf 633} (2006) 645
  %doi:10.1016/j.physletb.2005.12.034
  %[hep-lat/0507029].
  %%CITATION = doi:10.1016/j.physletb.2005.12.034;%%
  %102 citations counted in INSPIRE as of 25 Dec 2015
  

%\cite{D'Adda:2007ax}
%\bibitem{D'Adda:2007ax}
%  A.~D'Adda, I.~Kanamori, N.~Kawamoto and K.~Nagata,
  %``Exact Extended Supersymmetry on a Lattice: Twisted N=4 Super Yang-Mills in Three Dimensions,''
%  Nucl.\ Phys.\ B {\bf 798} (2008) 168
  %doi:10.1016/j.nuclphysb.2008.01.026
  %[arXiv:0707.3533 [hep-lat]].
  %%CITATION = doi:10.1016/j.nuclphysb.2008.01.026;%%
  %46 citations counted in INSPIRE as of 25 Dec 2015









% Catterall's geometrical approach

%\cite{Catterall_geometrical_approach}
\bibitem{Catterall_geometrical_approach}
  S.~Catterall, \ JHEP {\bf 0305} (2003) 038, \ JHEP {\bf 0411} (2004) 006, \ JHEP {\bf 0506} (2005) 027, \ JHEP {\bf 0603} (2006) 032, \ JHEP {\bf 0704} (2007) 015.
  
%\cite{Catterall:2003wd}
%\bibitem{Catterall:2003wd}
%  S.~Catterall,
  %``Lattice supersymmetry and topological field theory,''
%  JHEP {\bf 0305} (2003) 038
  %doi:10.1088/1126-6708/2003/05/038
  %[hep-lat/0301028].
  %%CITATION = doi:10.1088/1126-6708/2003/05/038;%%
  %91 citations counted in INSPIRE as of 25 Dec 2015
    

%\cite{Catterall:2004np}
%\bibitem{Catterall:2004np}
%  S.~Catterall,
  %``A Geometrical approach to N=2 super Yang-Mills theory on the two dimensional lattice,''
%  JHEP {\bf 0411} (2004) 006
  %doi:10.1088/1126-6708/2004/11/006
  %[hep-lat/0410052].
  %%CITATION = doi:10.1088/1126-6708/2004/11/006;%%
  %98 citations counted in INSPIRE as of 25 Dec 2015
  

%\cite{Catterall:2005fd}
%\bibitem{Catterall:2005fd}
%  S.~Catterall,
  %``Lattice formulation of N=4 super Yang-Mills theory,''
%  JHEP {\bf 0506} (2005) 027
  %doi:10.1088/1126-6708/2005/06/027
  %[hep-lat/0503036].
  %%CITATION = doi:10.1088/1126-6708/2005/06/027;%%
  %78 citations counted in INSPIRE as of 25 Dec 2015


    
%\cite{Catterall:2006jw}
%\bibitem{Catterall:2006jw}
%  S.~Catterall,
  %``Simulations of N=2 super Yang-Mills theory in two dimensions,''
%  JHEP {\bf 0603} (2006) 032
  %doi:10.1088/1126-6708/2006/03/032
  %[hep-lat/0602004].
  %%CITATION = doi:10.1088/1126-6708/2006/03/032;%%
  %33 citations counted in INSPIRE as of 25 Dec 2015



%\cite{Catterall:2006is}
%\bibitem{Catterall:2006is}
%  S.~Catterall,
  %``On the restoration of supersymmetry in twisted two-dimensional lattice Yang-Mills theory,''
%  JHEP {\bf 0704} (2007) 015
  %doi:10.1088/1126-6708/2007/04/015
  %[hep-lat/0612008].
  %%CITATION = doi:10.1088/1126-6708/2007/04/015;%%
  %21 citations counted in INSPIRE as of 25 Dec 2015    
    



  
       
  





% Equivalence


%\cite{Unsal:2006qp}
\bibitem{Unsal:2006qp}
  M.~Unsal,
  %``Twisted supersymmetric gauge theories and orbifold lattices,''
  JHEP {\bf 0610} (2006) 089.
  %doi:10.1088/1126-6708/2006/10/089
  %[hep-th/0603046].
  %%CITATION = doi:10.1088/1126-6708/2006/10/089;%%
  %64 citations counted in INSPIRE as of 25 Dec 2015
  


%\cite{Takimi:2007nn}
\bibitem{Takimi:2007nn}
  T.~Takimi,
  %``Relationship between various supersymmetric lattice models,''
  JHEP {\bf 0707} (2007) 010.
  %doi:10.1088/1126-6708/2007/07/010
  %[arXiv:0705.3831 [hep-lat]].
  %%CITATION = doi:10.1088/1126-6708/2007/07/010;%%
  %36 citations counted in INSPIRE as of 25 Dec 2015



%\cite{Damgaard_Matsuura}
\bibitem{Damgaard_Matsuura}
P.~H.~Damgaard and S.~Matsuura,  JHEP {\bf 0708} (2007) 087, \ Phys.\ Lett.\ B {\bf 661} (2008) 52.


%\cite{Damgaard:2007xi}
%\bibitem{Damgaard:2007xi}
%  P.~H.~Damgaard and S.~Matsuura,
  %``Relations among Supersymmetric Lattice Gauge Theories via Orbifolding,''
%  JHEP {\bf 0708} (2007) 087
  %doi:10.1088/1126-6708/2007/08/087
  %[arXiv:0706.3007 [hep-lat]].
  %%CITATION = doi:10.1088/1126-6708/2007/08/087;%%
  %31 citations counted in INSPIRE as of 25 Dec 2015



%\cite{Damgaard:2008pa}
%\bibitem{Damgaard:2008pa}
%  P.~H.~Damgaard and S.~Matsuura,
  %``Geometry of Orbifolded Supersymmetric Lattice Gauge Theories,''
%  Phys.\ Lett.\ B {\bf 661} (2008) 52
  %doi:10.1016/j.physletb.2008.01.044
  %[arXiv:0801.2936 [hep-th]].
  %%CITATION = doi:10.1016/j.physletb.2008.01.044;%%
  %37 citations counted in INSPIRE as of 25 Dec 2015
      
      

%\cite{Damgaard:2007eh}
\bibitem{Damgaard:2007eh}
  P.~H.~Damgaard and S.~Matsuura,
  %``Lattice Supersymmetry: Equivalence between the Link Approach and Orbifolding,''
  JHEP {\bf 0709} (2007) 097.
  %doi:10.1088/1126-6708/2007/09/097
  %[arXiv:0708.4129 [hep-lat]].
  %%CITATION = doi:10.1088/1126-6708/2007/09/097;%%
  %34 citations counted in INSPIRE as of 25 Dec 2015
  
 
 
 

%\cite{Catterall:2007kn}
\bibitem{Catterall:2007kn}
  S.~Catterall,
  %``From Twisted Supersymmetry to Orbifold Lattices,''
  JHEP {\bf 0801} (2008) 048.
  %doi:10.1088/1126-6708/2008/01/048
  %[arXiv:0712.2532 [hep-th]].
  %%CITATION = doi:10.1088/1126-6708/2008/01/048;%%
  %52 citations counted in INSPIRE as of 25 Dec 2015
  
    

  
  
  


% Numerical simulations for susy breaking


%\cite{Kanamori_Sugino_Suzuki}
\bibitem{Kanamori_Sugino_Suzuki}
  I.~Kanamori, F.~Sugino and H.~Suzuki, \ Phys.\ Rev.\ D {\bf 77} (2008) 091502, \ Prog.\ Theor.\ Phys.\  {\bf 119} (2008) 797, \   I.~Kanamori, \ Phys.\ Rev.\ D {\bf 79} (2009) 115015.
 
%\cite{Kanamori:2007ye}
%\bibitem{Kanamori:2007ye}
%  I.~Kanamori, H.~Suzuki and F.~Sugino,
  %``Euclidean lattice simulation for dynamical supersymmetry breaking,''
%  Phys.\ Rev.\ D {\bf 77} (2008) 091502
  %doi:10.1103/PhysRevD.77.091502
  %[arXiv:0711.2099 [hep-lat]].
  %%CITATION = doi:10.1103/PhysRevD.77.091502;%%
  %35 citations counted in INSPIRE as of 25 Dec 2015


%\cite{Kanamori:2007yx}
%\bibitem{Kanamori:2007yx}
%  I.~Kanamori, F.~Sugino and H.~Suzuki,
  %``Observing dynamical supersymmetry breaking with euclidean lattice simulations,''
%  Prog.\ Theor.\ Phys.\  {\bf 119} (2008) 797
  %doi:10.1143/PTP.119.797
  %[arXiv:0711.2132 [hep-lat]].
  %%CITATION = doi:10.1143/PTP.119.797;%%
  %38 citations counted in INSPIRE as of 25 Dec 2015
    

%\cite{Kanamori:2009dk}
%\bibitem{Kanamori:2009dk}
%  I.~Kanamori,
  %``Vacuum energy of two-dimensional N=(2,2) super Yang-Mills theory,''
%  Phys.\ Rev.\ D {\bf 79} (2009) 115015
  %doi:10.1103/PhysRevD.79.115015
  %[arXiv:0902.2876 [hep-lat]].
  %%CITATION = doi:10.1103/PhysRevD.79.115015;%%
  %14 citations counted in INSPIRE as of 25 Dec 2015
    
  
  

%\cite{Catterall:2015tta}
\bibitem{Catterall:2015tta}
  S.~Catterall and A.~Veernala,
  %``Spontaneous supersymmetry breaking in two dimensional lattice super QCD,''
  JHEP {\bf 1510} (2015) 013.
  %doi:10.1007/JHEP10(2015)013
  %[arXiv:1505.00467 [hep-lat]].
  %%CITATION = doi:10.1007/JHEP10(2015)013;%%




%\cite{Beccaria:2004pa}
\bibitem{Beccaria:2004pa}
  M.~Beccaria, M.~Campostrini and A.~Feo,
  %``Supersymmetry breaking in two-dimensions: The Lattice N = 1 Wess-Zumino model,''
  Phys.\ Rev.\ D {\bf 69} (2004) 095010.
  %doi:10.1103/PhysRevD.69.095010
  %[hep-lat/0402007].
  %%CITATION = doi:10.1103/PhysRevD.69.095010;%%
  %26 citations counted in INSPIRE as of 25 Dec 2015



%\cite{Wozar:2011gu}
\bibitem{Wozar:2011gu}
  C.~Wozar and A.~Wipf,
  %``Supersymmetry Breaking in Low Dimensional Models,''
  Annals Phys.\  {\bf 327} (2012) 774.
  %doi:10.1016/j.aop.2011.11.015
  %[arXiv:1107.3324 [hep-lat]].
  %%CITATION = doi:10.1016/j.aop.2011.11.015;%%
  %18 citations counted in INSPIRE as of 25 Dec 2015


%\cite{Steinhauer:2014yaa}
\bibitem{Steinhauer:2014yaa}
  K.~Steinhauer and U.~Wenger,
  %``Spontaneous supersymmetry breaking in the 2D $\mathcal N=$1 Wess-Zumino model,''
  Phys.\ Rev.\ Lett.\  {\bf 113} (2014) 23,  231601.
  %doi:10.1103/PhysRevLett.113.231601
  %[arXiv:1410.6665 [hep-lat]].
  %%CITATION = doi:10.1103/PhysRevLett.113.231601;%%
  %3 citations counted in INSPIRE as of 25 Dec 2015



%\cite{Kawai:2010yj}
\bibitem{Kawai:2010yj}
  H.~Kawai and Y.~Kikukawa,
  %``A Lattice study of N=2 Landau-Ginzburg model using a Nicolai map,''
  Phys.\ Rev.\ D {\bf 83} (2011) 074502.
  %doi:10.1103/PhysRevD.83.074502
  %[arXiv:1005.4671 [hep-lat]].
  %%CITATION = doi:10.1103/PhysRevD.83.074502;%%
  %8 citations counted in INSPIRE as of 16 Mar 2016

  
        



%
% Gauge gravity duality
%

%\cite{Hanada_Nishimura}
\bibitem{Hanada_Nishimura}
  K.~N.~Anagnostopoulos, M.~Hanada, J.~Nishimura and S.~Takeuchi, \ Phys.\ Rev.\ Lett.\  {\bf 100} (2008) 021601, \ M.~Hanada, A.~Miwa, J.~Nishimura and S.~Takeuchi, \ Phys.\ Rev.\ Lett.\  {\bf 102} (2009) 181602, \ M.~Hanada, Y.~Hyakutake, J.~Nishimura and S.~Takeuchi, \ Phys.\ Rev.\ Lett.\  {\bf 102} (2009) 191602, \  M.~Hanada, J.~Nishimura, Y.~Sekino and T.~Yoneya, \ JHEP {\bf 1112} (2011) 020, \ M.~Hanada, Y.~Hyakutake, G.~Ishiki and J.~Nishimura, \ Science {\bf 344} (2014) 882.



%\cite{Anagnostopoulos:2007fw}
%\bibitem{Anagnostopoulos:2007fw}
%  K.~N.~Anagnostopoulos, M.~Hanada, J.~Nishimura and S.~Takeuchi,
  %``Monte Carlo studies of supersymmetric matrix quantum mechanics with sixteen supercharges at finite temperature,''
%  Phys.\ Rev.\ Lett.\  {\bf 100} (2008) 021601
  %doi:10.1103/PhysRevLett.100.021601
  %[arXiv:0707.4454 [hep-th]].
  %%CITATION = doi:10.1103/PhysRevLett.100.021601;%%
  %121 citations counted in INSPIRE as of 01 Mar 2016


%\cite{Hanada:2008gy}
%\bibitem{Hanada:2008gy}
%  M.~Hanada, A.~Miwa, J.~Nishimura and S.~Takeuchi,
  %``Schwarzschild radius from Monte Carlo calculation of the Wilson loop in supersymmetric matrix quantum mechanics,''
%  Phys.\ Rev.\ Lett.\  {\bf 102} (2009) 181602
  %doi:10.1103/PhysRevLett.102.181602
  %[arXiv:0811.2081 [hep-th]].
  %%CITATION = doi:10.1103/PhysRevLett.102.181602;%%
  %63 citations counted in INSPIRE as of 01 Mar 2016

%\cite{Hanada:2008ez}
%\bibitem{Hanada:2008ez}
%  M.~Hanada, Y.~Hyakutake, J.~Nishimura and S.~Takeuchi,
  %``Higher derivative corrections to black hole thermodynamics from supersymmetric matrix quantum mechanics,''
%  Phys.\ Rev.\ Lett.\  {\bf 102} (2009) 191602
  %doi:10.1103/PhysRevLett.102.191602
  %[arXiv:0811.3102 [hep-th]].
  %%CITATION = doi:10.1103/PhysRevLett.102.191602;%%
  %77 citations counted in INSPIRE as of 01 Mar 2016


%\cite{Hanada:2011fq}
%\bibitem{Hanada:2011fq}
%  M.~Hanada, J.~Nishimura, Y.~Sekino and T.~Yoneya,
  %``Direct test of the gauge-gravity correspondence for Matrix theory correlation functions,''
%  JHEP {\bf 1112} (2011) 020
  %doi:10.1007/JHEP12(2011)020
  %[arXiv:1108.5153 [hep-th]].
  %%CITATION = doi:10.1007/JHEP12(2011)020;%%
  %25 citations counted in INSPIRE as of 01 Mar 2016


%\cite{Hanada:2013rga}
%\bibitem{Hanada:2013rga}
%  M.~Hanada, Y.~Hyakutake, G.~Ishiki and J.~Nishimura,
  %``Holographic description of quantum black hole on a computer,''
%  Science {\bf 344} (2014) 882
  %doi:10.1126/science.1250122
  %[arXiv:1311.5607 [hep-th]].
  %%CITATION = doi:10.1126/science.1250122;%%
  %35 citations counted in INSPIRE as of 01 Mar 2016



%\cite{Catterall_Wiseman}
\bibitem{Catterall_Wiseman}
S.~Catterall and T.~Wiseman, \ JHEP {\bf 0712} (2007) 104, \ Phys.\ Rev.\ D {\bf 78} (2008) 041502, \ JHEP {\bf 1004} (2010) 077.



%\cite{Catterall:2007fp}
%\bibitem{Catterall:2007fp}
%  S.~Catterall and T.~Wiseman,
  %``Towards lattice simulation of the gauge theory duals to black holes and hot strings,''
%  JHEP {\bf 0712} (2007) 104
  %doi:10.1088/1126-6708/2007/12/104
  %[arXiv:0706.3518 [hep-lat]].
  %%CITATION = doi:10.1088/1126-6708/2007/12/104;%%
  %88 citations counted in INSPIRE as of 25 Dec 2015


%\cite{Catterall:2008yz}
%\bibitem{Catterall:2008yz}
%  S.~Catterall and T.~Wiseman,
  %``Black hole thermodynamics from simulations of lattice Yang-Mills theory,''
%  Phys.\ Rev.\ D {\bf 78} (2008) 041502
  %doi:10.1103/PhysRevD.78.041502
  %[arXiv:0803.4273 [hep-th]].
  %%CITATION = doi:10.1103/PhysRevD.78.041502;%%
  %93 citations counted in INSPIRE as of 25 Dec 2015


%\cite{Catterall:2009xn}
%\bibitem{Catterall:2009xn}
%  S.~Catterall and T.~Wiseman,
  %``Extracting black hole physics from the lattice,''
%  JHEP {\bf 1004} (2010) 077
  %doi:10.1007/JHEP04(2010)077
  %[arXiv:0909.4947 [hep-th]].
  %%CITATION = doi:10.1007/JHEP04(2010)077;%%
  %53 citations counted in INSPIRE as of 25 Dec 2015


%\cite{Kadoh:2015mka}
\bibitem{Kadoh:2015mka}
  D.~Kadoh and S.~Kamata,
  %``Gauge/gravity duality and lattice simulations of one dimensional SYM with sixteen supercharges,''
  arXiv:1503.08499 [hep-lat].
  %%CITATION = ARXIV:1503.08499;%%
  %6 citations counted in INSPIRE as of 25 Dec 2015



%\cite{Filev:2015hia}
\bibitem{Filev:2015hia}
  V.~G.~Filev and D.~O'Connor,
  %``The BFSS model on the lattice,''
  arXiv:1506.01366 [hep-th].
  %%CITATION = ARXIV:1506.01366;%%
  %3 citations counted in INSPIRE as of 25 Dec 2015






%\cite{Catterall:2010fx}
\bibitem{Catterall:2010fx}
  S.~Catterall, A.~Joseph and T.~Wiseman,
  %``Thermal phases of D1-branes on a circle from lattice super Yang-Mills,''
  JHEP {\bf 1012} (2010) 022.
  %doi:10.1007/JHEP12(2010)022
  %[arXiv:1008.4964 [hep-th]].
  %%CITATION = doi:10.1007/JHEP12(2010)022;%%
  %40 citations counted in INSPIRE as of 25 Dec 2015




%\cite{Giguere:2015cga}
\bibitem{Giguere:2015cga}
  E.~Giguere and D.~Kadoh,
  %``Restoration of supersymmetry in two-dimensional SYM with sixteen supercharges on the lattice,''
  JHEP {\bf 1505} (2015) 082.
  %doi:10.1007/JHEP05(2015)082
  %[arXiv:1503.04416 [hep-lat]].
  %%CITATION = doi:10.1007/JHEP05(2015)082;%%
  %2 citations counted in INSPIRE as of 25 Dec 2015
  
 
 
%\cite{Giguere:2016}
\bibitem{Giguere:2016}
  E.~Giguere and D.~Kadoh, 
  in preparation.
 

%\cite{Catterall:2011pd,  Catterall:2012yq, Catterall:2013roa, Catterall:2014vka, \cite{Catterall:2014mha}


%\cite{Catterall:2011pd}
\bibitem{Catterall:2011pd}
  S.~Catterall, E.~Dzienkowski, J.~Giedt, A.~Joseph and R.~Wells,
  %``Perturbative renormalization of lattice N=4 super Yang-Mills theory,''
  JHEP {\bf 1104} (2011) 074.
  %doi:10.1007/JHEP04(2011)074
  %[arXiv:1102.1725 [hep-th]].
  %%CITATION = doi:10.1007/JHEP04(2011)074;%%
  %37 citations counted in INSPIRE as of 25 Dec 2015



%\cite{Catterall:2012yq}
\bibitem{Catterall:2012yq}
  S.~Catterall, P.~H.~Damgaard, T.~Degrand, R.~Galvez and D.~Mehta,
  %``Phase Structure of Lattice N=4 Super Yang-Mills,''
  JHEP {\bf 1211} (2012) 072.
  %doi:10.1007/JHEP11(2012)072
  %[arXiv:1209.5285 [hep-lat]].
  %%CITATION = doi:10.1007/JHEP11(2012)072;%%
  %18 citations counted in INSPIRE as of 25 Dec 2015  

  
%\cite{Catterall:2013roa}
\bibitem{Catterall:2013roa}
  S.~Catterall, J.~Giedt and A.~Joseph,
  %``Twisted supersymmetries in lattice ${\cal N}=4$ super Yang-Mills theory,''
  JHEP {\bf 1310} (2013) 166.
  %doi:10.1007/JHEP10(2013)166
  %[arXiv:1306.3891 [hep-lat]].
  %%CITATION = doi:10.1007/JHEP10(2013)166;%%
  %16 citations counted in INSPIRE as of 25 Dec 2015

    

%\cite{Catterall:2014vka}
\bibitem{Catterall:2014vka}
  S.~Catterall, D.~Schaich, P.~H.~Damgaard, T.~DeGrand and J.~Giedt,
  %``N=4 Supersymmetry on a Space-Time Lattice,''
  Phys.\ Rev.\ D {\bf 90} (2014) 6,  065013.
  %doi:10.1103/PhysRevD.90.065013
  %[arXiv:1405.0644 [hep-lat]].
  %%CITATION = doi:10.1103/PhysRevD.90.065013;%%
  %14 citations counted in INSPIRE as of 25 Dec 2015



%\cite{Catterall:2014mha}
\bibitem{Catterall:2014mha}
  S.~Catterall and J.~Giedt,
  %``Real space renormalization group for twisted lattice $ \mathcal{N} $ =4 super Yang-Mills,''
  JHEP {\bf 1411} (2014) 050.
  %doi:10.1007/JHEP11(2014)050
  %[arXiv:1408.7067 [hep-lat]].
  %%CITATION = doi:10.1007/JHEP11(2014)050;%%
  %9 citations counted in INSPIRE as of 25 Dec 2015


  
 %\cite{Klebanov:1996un}
\bibitem{Klebanov:1996un}
  I.~R.~Klebanov and A.~A.~Tseytlin,
  %``Entropy of near extremal black p-branes,''
  Nucl.\ Phys.\ B {\bf 475} (1996) 164.
  %doi:10.1016/0550-3213(96)00295-7
  %[hep-th/9604089].
  %%CITATION = doi:10.1016/0550-3213(96)00295-7;%%
  %325 citations counted in INSPIRE as of 01 Mar 2016







%
%  Other interesting results 
%


%\cite{Suzuki:2012wx}
\bibitem{Suzuki:2012wx}
  H.~Suzuki,
  %``Remark on the energy-momentum tensor in the lattice formulation of 4D $\mathcal{N}=1$ SYM,''
  Phys.\ Lett.\ B {\bf 719} (2013) 435.
  %doi:10.1016/j.physletb.2013.01.028
  %[arXiv:1209.5155 [hep-lat]].
  %%CITATION = doi:10.1016/j.physletb.2013.01.028;%%
  %5 citations counted in INSPIRE as of 25 Dec 2015


%\cite{Suzuki:2013gi}
\bibitem{Suzuki:2013gi}
  H.~Suzuki,
  %``Ferrara--Zumino supermultiplet and the energy-momentum tensor in the lattice formulation of 4D $\mathcal{N}=1$ SYM,''
  Nucl.\ Phys.\ B {\bf 868} (2013) 459.
  %doi:10.1016/j.nuclphysb.2012.11.023
  %[arXiv:1209.2473 [hep-lat]].
  %%CITATION = doi:10.1016/j.nuclphysb.2012.11.023;%%
  %4 citations counted in INSPIRE as of 25 Dec 2015




% CLR

%\cite{Kato:2013sba}
\bibitem{Kato:2013sba}
  M.~Kato, M.~Sakamoto and H.~So,
  %``A criterion for lattice supersymmetry: cyclic Leibniz rule,''
  JHEP {\bf 1305} (2013) 089.
  %doi:10.1007/JHEP05(2013)089
  %[arXiv:1303.4472 [hep-lat]].
  %%CITATION = doi:10.1007/JHEP05(2013)089;%%
  %4 citations counted in INSPIRE as of 25 Dec 2015


%\cite{Kadoh:2015zza}
\bibitem{Kadoh:2015zza}
  D.~Kadoh and N.~Ukita,
  %``General solution of the cyclic Leibniz rule,''
  PTEP {\bf 2015} (2015) 10,  103B04.
  %doi:10.1093/ptep/ptv140
  %[arXiv:1503.06922 [hep-lat]].
  %%CITATION = doi:10.1093/ptep/ptv140;%%




%\cite{Kikuchi:2014rla}
%\bibitem{Kikuchi:2014rla}
%  K.~Kikuchi and T.~Onogi,
  %``Generalized Gradient Flow Equation and Its Application to Super Yang-Mills Theory,''
%  JHEP {\bf 1411} (2014) 094
  %doi:10.1007/JHEP11(2014)094
  %[arXiv:1408.2185 [hep-th]].
  %%CITATION = doi:10.1007/JHEP11(2014)094;%%
  %8 citations counted in INSPIRE as of 25 Dec 2015


%\cite{Matsuura_Misumi_Ohta}
\bibitem{Matsuura_Misumi_Ohta}
  S.~Matsuura, T.~Misumi and K.~Ohta, \ PTEP {\bf 2014} (2014) 12,  123B01, \ PTEP {\bf 2015} (2015) 3,  033B07.
  
  
%\cite{Matsuura:2014kha}
%\bibitem{Matsuura:2014kha}
%  S.~Matsuura, T.~Misumi and K.~Ohta,
  %``Topologically twisted N = (2, 2) supersymmetric Yang–Mills theory on an arbitrary discretized Riemann surface,''
%  PTEP {\bf 2014} (2014) 12,  123B01
  %doi:10.1093/ptep/ptu153
  %[arXiv:1408.6998 [hep-lat]].
  %%CITATION = doi:10.1093/ptep/ptu153;%%
  %1 citations counted in INSPIRE as of 25 Dec 2015


%\cite{Matsuura:2014nga}
%\bibitem{Matsuura:2014nga}
%  S.~Matsuura, T.~Misumi and K.~Ohta,
  %``Exact Results in Discretized Gauge Theories,''
%  PTEP {\bf 2015} (2015) 3,  033B07
  %doi:10.1093/ptep/ptv021
  %[arXiv:1411.4466 [hep-th]].
  %%CITATION = doi:10.1093/ptep/ptv021;%%




% Loop formulation

%\cite{Steinhauer:2014oda}
%\bibitem{Steinhauer:2014oda}
%  K.~Steinhauer and U.~Wenger,
  %``Loop formulation of supersymmetric Yang-Mills quantum mechanics,''
%  JHEP {\bf 1412} (2014) 044
  %doi:10.1007/JHEP12(2014)044
  %[arXiv:1410.0235 [hep-lat]].
  %%CITATION = doi:10.1007/JHEP12(2014)044;%%
  %5 citations counted in INSPIRE as of 25 Dec 2015


%\cite{Baumgartner:2014nka}
%\bibitem{Baumgartner:2014nka}
%  D.~Baumgartner and U.~Wenger,
  %``Supersymmetric quantum mechanics on the lattice: I. Loop formulation,''
%  Nucl.\ Phys.\ B {\bf 894} (2015) 223
  %doi:10.1016/j.nuclphysb.2015.03.001
  %[arXiv:1412.5393 [hep-lat]].
  %%CITATION = doi:10.1016/j.nuclphysb.2015.03.001;%%
  %4 citations counted in INSPIRE as of 25 Dec 2015


%\cite{Baumgartner:2015qba}
%\bibitem{Baumgartner:2015qba}
%  D.~Baumgartner and U.~Wenger,
  %``Supersymmetric quantum mechanics on the lattice: II. Exact results,''
%  Nucl.\ Phys.\ B {\bf 897} (2015) 39
  %doi:10.1016/j.nuclphysb.2015.05.010
  %[arXiv:1503.05232 [hep-lat]].
  %%CITATION = doi:10.1016/j.nuclphysb.2015.05.010;%%
  %2 citations counted in INSPIRE as of 25 Dec 2015


%\cite{Baumgartner:2015zna}
%\bibitem{Baumgartner:2015zna}
%  D.~Baumgartner and U.~Wenger,
  %``Supersymmetric quantum mechanics on the lattice: III. Simulations and algorithms,''
%  Nucl.\ Phys.\ B {\bf 899} (2015) 375
  %doi:10.1016/j.nuclphysb.2015.07.020
  %[arXiv:1505.07397 [hep-lat]].
  %%CITATION = doi:10.1016/j.nuclphysb.2015.07.020;%%
  %2 citations counted in INSPIRE as of 25 Dec 2015





% The other methods


%\cite{Harada:2003bs}
%\bibitem{Harada:2003bs}
%  M.~Harada and S.~Pinsky,
  %``N=(1,1) superYang-Mills on a (2+1)-dimensional transverse lattice with one exact supersymmetry,''
%  Phys.\ Lett.\ B {\bf 567} (2003) 277
  %doi:10.1016/j.physletb.2003.06.035
  %[hep-lat/0303027].
  %%CITATION = doi:10.1016/j.physletb.2003.06.035;%%
  %25 citations counted in INSPIRE as of 25 Dec 2015
  



%\cite{Suzuki:2005dx}
%\bibitem{Suzuki:2005dx}
%  H.~Suzuki and Y.~Taniguchi,
  %``Two-dimensional N = (2,2) super Yang-Mills theory on the lattice via dimensional reduction,''
%  JHEP {\bf 0510} (2005) 082
  %doi:10.1088/1126-6708/2005/10/082
  %[hep-lat/0507019].
  %%CITATION = doi:10.1088/1126-6708/2005/10/082;%%
  %40 citations counted in INSPIRE as of 25 Dec 2015



%\cite{Elliott:2005bd}
%\bibitem{Elliott:2005bd}
%  J.~W.~Elliott and G.~D.~Moore,
  %``Three dimensional N=2 supersymmetry on the lattice,''
%  JHEP {\bf 0511} (2005) 010
  %doi:10.1088/1126-6708/2005/11/010
  %[hep-lat/0509032].
  %%CITATION = doi:10.1088/1126-6708/2005/11/010;%%
  %28 citations counted in INSPIRE as of 25 Dec 2015







% No go theorems and CLR


%\cite{Kato:2008sp}
\bibitem{Kato:2008sp}
  M.~Kato, M.~Sakamoto and H.~So,
  %``Taming the Leibniz Rule on the Lattice,''
  JHEP {\bf 0805} (2008) 057.
  %doi:10.1088/1126-6708/2008/05/057
  %[arXiv:0803.3121 [hep-lat]].
  %%CITATION = doi:10.1088/1126-6708/2008/05/057;%%
  %34 citations counted in INSPIRE as of 29 Feb 2016


%\cite{Bergner:2009vg}
\bibitem{Bergner:2009vg}
  G.~Bergner,
  %``Complete supersymmetry on the lattice and a No-Go theorem,''
  JHEP {\bf 1001} (2010) 024.
  %doi:10.1007/JHEP01(2010)024
  %[arXiv:0909.4791 [hep-lat]].
  %%CITATION = doi:10.1007/JHEP01(2010)024;%%
  %31 citations counted in INSPIRE as of 25 Dec 2015


  

%\cite{Hanada:2010gs}
%\bibitem{Hanada:2010gs}
%  M.~Hanada,
  %``A proposal of a fine tuning free formulation of 4d N = 4 super Yang-Mills,''
%  JHEP {\bf 1011} (2010) 112
  %doi:10.1007/JHEP11(2010)112
  %[arXiv:1009.0901 [hep-lat]].
  %%CITATION = doi:10.1007/JHEP11(2010)112;%%
  %35 citations counted in INSPIRE as of 25 Dec 2015
    
      

%\cite{Hanada_Matsuura_Sugino}
\bibitem{Hanada_Matsuura_Sugino}
  M.~Hanada, S.~Matsuura and F.~Sugino, \ Prog.\ Theor.\ Phys.\  {\bf 126} (2011) 597, \ Nucl.\ Phys.\ B {\bf 857} (2012) 335, \ S.~Matsuura and F.~Sugino, \ arXiv:1508.00707 [hep-th].
  
  
%\cite{Hanada:2010kt}
%\bibitem{Hanada:2010kt}
%  M.~Hanada, S.~Matsuura and F.~Sugino,
  %``Two-dimensional lattice for four-dimensional N=4 supersymmetric Yang-Mills,''
%  Prog.\ Theor.\ Phys.\  {\bf 126} (2011) 597
  %doi:10.1143/PTP.126.597
  %[arXiv:1004.5513 [hep-lat]].
  %%CITATION = doi:10.1143/PTP.126.597;%%
  %48 citations counted in INSPIRE as of 25 Dec 2015 



%\cite{Hanada:2011qx}
%\bibitem{Hanada:2011qx}
%  M.~Hanada, S.~Matsuura and F.~Sugino,
  %``Non-perturbative construction of 2D and 4D supersymmetric Yang-Mills theories with 8 supercharges,''
%  Nucl.\ Phys.\ B {\bf 857} (2012) 335
  %doi:10.1016/j.nuclphysb.2011.12.014
  %[arXiv:1109.6807 [hep-lat]].
  %%CITATION = doi:10.1016/j.nuclphysb.2011.12.014;%%
  %20 citations counted in INSPIRE as of 25 Dec 2015



%\cite{Matsuura:2015caa}
%\bibitem{Matsuura:2015caa}
%  S.~Matsuura and F.~Sugino,
  %``A one-loop test for construction of 4D N=4 SYM from 2D SYM via fuzzy sphere geometry,''
%  arXiv:1508.00707 [hep-th].
  %%CITATION = ARXIV:1508.00707;%%





 

% Review


%\cite{Catterall:2009it}
\bibitem{Catterall:2009it}
  S.~Catterall, D.~B.~Kaplan and M.~Unsal,
  %``Exact lattice supersymmetry,''
  Phys.\ Rept.\  {\bf 484} (2009) 71.
  %doi:10.1016/j.physrep.2009.09.001
  %[arXiv:0903.4881 [hep-lat]].
  %%CITATION = doi:10.1016/j.physrep.2009.09.001;%%
  %108 citations counted in INSPIRE as of 25 Dec 2015












\end{thebibliography}
\end{document}